\documentclass[useAMS,usenatbib,usegraphicx]{mn2e}
\usepackage{amsmath}

\graphicspath{{./figures/}}

\DeclareMathAlphabet{\mathpzc}{OT1}{pzc}{m}{it}

\newcommand{\pd}{\partial}           

\newcommand{\Uvec}[1]%
{\ensuremath{\mathbf{e}_{#1}}}  

\renewcommand{\vec}[1]%
{\ensuremath{\bmath{#1}}}       

\newcommand{\B}{\vec{B}}        
\newcommand{\E}{\vec{E}}        

\newcommand{\GJ}[1]{\ensuremath{#1_\textrm{\tiny GJ}}}    
\newcommand{\GJNS}[1]{\ensuremath{#1_\textrm{\tiny GJ}^0}}    
 
\newcommand{\Pol}[1]{\ensuremath{#1_\mathrm{pol}}}    
\newcommand{\PC}[1]{\ensuremath{#1_\mathrm{pc}}}

\newcommand{\OmF}{\ensuremath{\Omega_\mathrm{F}}}

\newcommand{\PsiL}{\ensuremath{\Psi_\mathrm{pc}}} 
\newcommand{\PsiM}{\ensuremath{\Psi_\mathrm{M}}} 

\newcommand{\Wmd}{\protect{\ensuremath{W_\mathrm{md}}}}

\newcommand{\RLC}{\ensuremath{R_\mathrm{LC}}}     
\newcommand{\RNS}{\ensuremath{R_\mathrm{NS}}}

\newcommand{\PsiEx}{\ensuremath{\psi_\textrm{ex}}}

\newcommand{\DVba}{\ensuremath{\Delta{}V_\textrm{10}}}
\newcommand{\DVbe}{\ensuremath{\Delta{}V_\textrm{1e}}}
\newcommand{\DVae}{\ensuremath{\Delta{}V_\textrm{0e}}}

\newcommand{\Max}[1]{\ensuremath{#1_\textrm{max}}}    
\newcommand{\Min}[1]{\ensuremath{#1_\textrm{min}}}    

\newcommand{\Const}[1]{\ensuremath{#1^\textrm{c}}}    
\newcommand{\Opt}[1]{\ensuremath{#1^\textrm{s}}}

\newcommand{\PapI}{Paper~I}
\newcommand{\PapII}{Paper~II}

\title[Differentially rotating force-free magnetosphere]%
{Differentially rotating force-free magnetosphere of an aligned
  rotator: analytical solutions in split-monopole approximation}
\author[A.~N.~Timokhin]{%
  A.~N.~Timokhin\thanks{E-mail: atim@sai.msu.ru}\\
  Physics Department, Ben-Gurion University of the Negev, POB  653, 84105 Beer-Sheva, Israel\\
  Sternberg Astronomical Institute, Universitetskij pr. 13, 119992
  Mocsow, Russia}

\begin{document}

\date{Accepted 2007 April 16, Received 2007 February 27; in original form 2007 February 27}

\pagerange{\pageref{firstpage}--\pageref{lastpage}} \pubyear{}

\maketitle

\label{firstpage}

\begin{abstract}
  In this paper we consider stationary force-free magnetosphere of an
  aligned rotator when plasma in the open field line region rotates
  differentially due to presence of a zone with the accelerating
  electric field in the polar cap of pulsar.  We study the impact of
  differential rotation on the current density distribution in the
  magnetosphere.  Using split-monopole approximation we obtain
  analytical expressions for physical parameters of differentially
  rotating magnetosphere.  We find the range of admitted current
  density distributions under the requirement that the potential drop
  in the polar cap is less than the vacuum potential drop.  We show
  that the current density distribution could deviate significantly
  from the ``classical'' Michel distribution and could be made almost
  constant over the polar cap even when the potential drop in the
  accelerating zone is of the order of 10 per cents of the vacuum
  potential drop.  We argue that differential rotation of the open
  magnetic field lines could play an important role in adjusting
  between the magnetosphere and the polar cap cascade zone and could
  affect the value of pulsar breaking index.
\end{abstract}


\section{Introduction}
\label{sec:introduction}

The physics of radiopulsars is still not fully understood despite
large effort of many theoreticians in this field.  It is generally
assumed that radiopulsar has MHD-like magnetosphere which is very
close to the force-free stage -- the model firstly introduced by
\citet{GJ}.  For many years solution of force-free MHD equations was
the problem, even for the simplest case of an aligned pulsar.  Now to
solve the Grade-Shafranov equation describing structure of a
force-free magnetosphere of an aligned pulsar is not a problem anymore
\citep[see e.g.][]{CKF,Goodwin/04,Gruzinov:PSR,Timokhin2006:MNRAS1}.
Stationary magnetosphere configurations for an aligned rotator were
obtained also a the final stage in non-stationary numerical modelling
\citep{Komissarov06,McKinney:NS:06,Bucciantini06,Spitkovsky:incl:06}.
Even the case of an inclined rotator was studied numerically
\citep{Spitkovsky:incl:06}.  However, the pulsar magnetosphere is a
very complicated physical system because most of the current carriers
(electrons and positrons) are produced inside the system, in the polar
cap cascades.  Production of electron-positron pairs is a process with
a threshold, so it could operate only under specific conditions and,
generally speaking, not any given current density could flow through
the cascade zone.

In magnetohydrodynamics (MHD) the current density distribution is not
a free ``parameter'', it is obtained in course of solving of MHD
equations.  In case of pulsars obtaining a solution of MHD equations
does not solve the problem, because it could happen that the polar cap
cascade zone could not provide the required current density
distribution and, hence, support the particular configurations of the
magnetosphere.  In terms of MHD the polar cap cascade zone sets
complicated boundary conditions at the foot points of the open
magnetic field lines and any self-consistent solution of the problem
must match them.  The most ``natural'' configuration of the
magnetosphere of an aligned rotator, when the last closed field line
extends up to the light cylinder, requires current density
distribution which could not be supported by stationary
electromagnetic cascades in the polar cap of pulsar
\citep[see][hereafter \PapI]{Timokhin2006:MNRAS1}.  That configuration
requires that in some parts of the polar cap the electric current
flows against the preferable direction of the accelerating electric
field.  This seems to be not possible also for non-stationary
cascades, although this problem requires more carefully investigation
than it has been done before
\citep{Fawley/PhDT:1978,AlBer/Krotova:1975,Levinson05}.  So, the
structure of the magnetosphere should be different from this simple
picture.  The magnetosphere of a pulsar would have a configuration
with the current density distribution which can flow through the
polar cap cascade zone without suppression of electron-positron pair
creation.  Whether such configuration exists is still an open
question, i.e. a possibility that the real pulsar magnetosphere has
large domains where MHD approximation is broken could not be
completely excluded too \citep[see e.g.][]{Arons79,MichelBook}.

As the pulsar magnetosphere and the polar cap cascade zone have too
different characteristic timescales, it would be barely possible to
proceed with modelling of the whole system at once.  Therefore, these
physical systems should be modelled separately and the whole set of
solutions for each system should be found, in order to find compatible
ones.  Namely, we suggest the following approach to the construction
of the pulsar magnetosphere model: one should find out which currents
could flow through the force-free pulsar magnetosphere and compare
them with the currents being able to flow through the polar cap
cascade zone.  In this work we deal with the first part of the the
suggested ``program''. Namely, we consider the range of possible
current density distributions in force-free magnetosphere of an
aligned rotator.

Force-free magnetosphere of an aligned rotator is the simplest
possible case of an MHD-like pulsar magnetosphere and needs to be
investigated in the first place.  This system has two physical degrees
of freedoms i) the size of the closed field line zone, and ii) the
distribution of the angular velocity of open magnetic field lines.  In
each stationary configuration the current density distribution is
fixed.  Considering different configurations by changing (i) and (ii)
and keeping them in reasonable range the whole set of admitted current
density distributions can be found.  Differential rotation of the open
field lines is caused by variation of the accelerating electric
potential in the cascade zone across the polar cap.  Theories of
stationary polar cap cascades predict rather small potential drop and
in this case only one degree of freedom is left -- the size of the
zone with closed magnetic field lines.  This case was studied in
details in \PapI, with the results that stationary polar cap cascades
are incompatible with stationary force-free magnetosphere.  So, most
probably the polar cap cascades operate in non-stationary regime.  For
non-stationary cascades the average potential drop in the accelerating
zone could be larger than the drop maintained by stationary cascades.
Hence, the open magnetic field lines may rotate with significantly
different angular velocities even in magnetospheres of young
pulsars.  On the other hand, for old pulsars the potential drop in the
cascade zone is large, and magnetospheres of such pulsars should
rotate essentially differentially anyway.

The case of differentially rotating pulsar magnetosphere was not
investigated in details before Although some authors addressed the
case when the open magnetics field lines rotate differently than the
NS, but only the case of constant angular velocity was considered
\citep[e.g.][]{GurevichBeskinIstomin_Book,Contopoulos05}.  The first
attempt to construct a self-consistent model of pulsar magnetosphere
with \emph{differentially} rotating open field line zone was made in
\citet{Timokhin::PSREQ2/2007}, hereafter \PapII.  In that paper we
considered only the case when the angular velocity of the open field
lines is less than the angular velocity of the NS. We have shown that
the current density can be made almost constant over the polar cap,
although on a cost of a large potential drop in the accelerating zone.
The angular velocity distributions was chosen ad hoc and the analysis
of the admitted range for current density distributions was not
performed.

In this paper we discuss properties of differentially rotating
magnetosphere of an aligned rotator in general and elaborate the
limits on the differential rotation.  We study in detail the case when
the current density in the polar cap is a linear function on the
magnetic flux. It allows us to obtain main relations analytically.  We
find the range in which physical parameters of the magnetosphere could
vary, requiring that a) the potential drop in the polar cap is not
greater that the vacuum potential drop and b) the current in the polar
cap does not change its direction.

The plan of the paper is as follows. In Section~\ref{sec:basic-model}
we discuss basic properties of differentially rotating force-free
magnetosphere of an aligned rotator and derive equations for angular
velocity distribution, current density and the Goldreich-Julian charge
density in the magnetosphere. In Section~\ref{sec:Equation_for_V} we
derive equations for the potential drop which supports configurations
with linear current density distribution in the polar cap of pulsar
and give their general solutions.  In Section~\ref{sec:main-results}
we analyse the physical properties of admitted magnetosphere
configurations: the current density distribution, the maximum
potential drop, the angular velocity of the open magnetic field lines,
the Goldreich-Julian current density, the spindown rate and the total
energy of the magnetosphere.  At the end of that section we we consider
as examples two sets of solutions: the one with constant current
densities and the another one with the smallest potential drops. In
Section~\ref{sec:Discussion} we summarise the results, discuss
limitation of the used approximation and briefly describe possible
modification of the obtained solutions which will arise in truly
self-consistent model.  In that section we also discuss the issue with
the pulsar braking index.

\section{Differentially rotating magnetosphere: basic properties}
\label{sec:basic-model}

\subsection{Pulsar equation}
\label{sec:general-equation}

Here as in Papers~I,II we consider magnetosphere of an aligned rotator
that is at the coordinate origin and has dipolar magnetic field.  We
use normalisations similar%
\footnote{note that here in contrast to \PapI{} $\Psi$ is already
  dimensionless}
to the ones in \PapI, but now we write all equations in the spherical
coordinates $(r,\theta,\phi)$.  We normalise all distances to the
light cylinder radius of the corotating magnetosphere
$\RLC\equiv{}c/\Omega$, where $\Omega$ is the angular velocity of the
neutron star (NS), $c$ is the speed of light.  For the considered
axisymmetric case the magnetic field can be expressed through two
dimensionless scalar functions $\Psi$ and $S$ as (cf.  eq.~(8) in
\PapI)
\begin{equation}
  \label{eq:B}
  \B = \frac{\mu}{\RLC^3}
    \frac{ \vec{\nabla}\Psi \times \vec{e_\phi} + S\vec{e_\phi} }{r\sin\theta}
  \,,
\end{equation}
where $\vec{e_\phi}$ is the unit azimuthal, toroidal vector.
$\mu=B_0\RNS^3/2$ is the magnetic moment of the NS; $B_0$ is the
magnetic field strength at the magnetic pole, $\RNS$ is the NS radius.
The scalar function $\Psi$ is related to the magnetic flux as
$\Phi_\mathrm{mag}(\varpi,Z)=2\pi\,(\mu/\RLC)\:\Psi(r,\theta)$.
$\Phi_\mathrm{mag}$ is the magnetic flux trough a circle of a radius
$\varpi=r\sin\theta$ with its centre at the point on the rotation axis
being at the distance $Z=r\cos\theta$ from the NS.  The lines of
constant $\Psi$ coincides with magnetic field lines.  The scalar
function $S$ is related to the total current $J$ \emph{outflowing}
trough the same circle by
$J(\varpi,Z)=1/2\,(\mu/\RLC^2)\,c\:S(r,\theta)$.

The electric field in the force-free magnetosphere is given by
\begin{equation}
  \label{eq:E}
  \E = - \frac{\mu}{\RLC^3}\: \beta\, \nabla \Psi
  \, ,
\end{equation}
where $\beta$ is the ratio of the angular velocity of the magnetic
field lines rotation $\OmF$ to the angular velocity of the NS,
$\beta\equiv\OmF/\Omega$ (cf. eq.~(14) in \PapI).  The difference of
the angular velocity of a magnetic field line from the angular
velocity of the NS is due to potential drop along that line in the
polar cap acceleration zone.

For these dimensionless functions the equation describing the
stationary force-free magnetosphere, the so-called pulsar equation
\citep{Michel73:b,ScharlemannWagoner73,Okamoto74}, takes the form (cf.
eq.~20 in \PapI)
\begin{multline}
  \label{eq:PsrEq}
  \left[ 1 - (\beta r \sin\theta)^2 \right] \Delta\Psi -
  \frac{2}{r}
  \left( 
    \pd_r \Psi + \frac{\cos\theta}{\sin\theta}\frac{\pd_\theta\Psi}{r} 
  \right) +
  \\
  + S \frac{d S}{d \Psi}
  - \beta \frac{d \beta}{d \Psi} \left(r \sin\theta\: \nabla \Psi \right)^2
  =  0
  \,.
\end{multline}
This equation express the force balance across the magnetic field
lines.  At the light cylinder the coefficient by the Laplacian
goes to zero and the pulsar equation reduces to
\begin{equation}
  \label{eq:CondAtLC}
  S \frac{d S}{d \Psi} =
  \frac{1}{\beta} \frac{d \beta}{d \Psi} \left( \nabla \Psi \right)^2 +
  2\beta \sin\theta\, 
  \left( 
        \pd_r \Psi + \beta \cos\theta\, \pd_\theta\Psi 
  \right)
  \,.
\end{equation}
Each smooth solution must satisfy these two equations and the problem
of a solving the pulsar equations transforms to an eigenfunction
problem for the poloidal current function $S$ (see e.g.  Section~2.3
in \PapI).  Equation~(\ref{eq:CondAtLC}) could also be considered as
an equation on the poloidal current.

\begin{figure}
  \includegraphics[clip,width=\columnwidth]{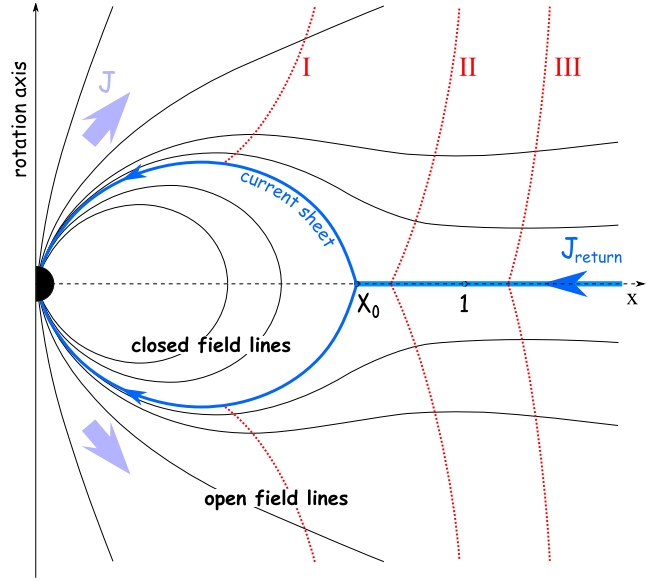}
  \caption{Structure of the magnetosphere of an aligned rotator
    (schematic picture).  Magnetic field lines are shown by solid
    lines.  Outflowing current $J$ along open magnetic field lines and
    returning current $J_\mathrm{return}$ in the current sheet,
    separating the open and closed magnetic field line zones, are
    indicated by arrows.  The current sheet is along the last open
    magnetic field line, corresponding to the value of the flux
    function $\PsiL$.  Distances are measured in units of the light
    cylinder radius for the corotating magnetosphere $\RLC$, i.e.  the
    point with $x=1$ marks the position of the light cylinder in the
    corotating magnetosphere.  The null point $x_0$ could lie anywhere
    inside the interval $[0,1]$.  Possible positions of the real light
    cylinder are shown by dotted lines.  The line \textbf{I}
    corresponds to the case when $1/\beta(\PsiL)<x_0$; \textbf{II} --
    to $x_0<1/\beta(\PsiL)<1$; \textbf{III} -- to $1<1/\beta(\PsiL)$
    (see the text further in the article). }
  \label{fig:Magnetosphere}
\end{figure}

We adopt for the magnetosphere the configuration with the so-called
Y-null point.  Namely, we assume that the magnetosphere is divided in
two zones, the first one with closed magnetic field lines, which
extend from the NS up to the neutral point being at distance $x_0$
from the NS, and the second one, where magnetic field lines are open
and extend to infinity (see Fig.~\ref{fig:Magnetosphere}).  In the
closed magnetic field line zone plasma corotates with the NS, there is
no poloidal current along field lines, and the magnetic field lines
there are equipotential.  Apparently this zone can not extend beyond
the light cylinder.  In the the rest of the magnetosphere magnetic
field lines are open due to poloidal current produced by outflowing
charged particles.  The return current, needed to keep the NS charge
neutral flows in a thin region (current sheet) along the equatorial
plane and then along the last open magnetic field line.  We assume
that this picture is stationary on the time scale of the order of the
period of the NS rotation.  As it was outlined in \PapI{}, the polar
cap cascades in pulsars are most probably non-stationary.  The
characteristic time scale of the polar cap cascades
$\sim{}h/c\sim3\cdot10^{-5}$~sec ($h$ is the length of the
acceleration zone being of the order of $\RNS$) is much shorter that
the pulsar period (for most pulsars being $\gg{}10^{-3}$~sec).  So,
for the global magnetosphere structure only time average of the
physical parameters connected to the cascade zone are important.  In
the rest of the paper, when we discuss physical parameters set by the
cascade zone we will always mean the \emph{average} values of them,
unless the opposite is explicitly stated.

Differential rotation of the open magnetic field lines which is caused
by presence of a zone with the accelerating electric field in the
polar cap of pulsar i) contributes to the force balance across
magnetic field lines (the last term in eq.(\ref{eq:PsrEq})), ii)
modifies the current density in the magnetosphere (the first term in
r.h.s. of eq.~(\ref{eq:CondAtLC})), and iii) changes the position of
the light cylinder, where condition~(\ref{eq:CondAtLC}) must be
satisfied.  Note that for the case i) and ii) the derivative
$d\beta/d\Psi$, i.e. the form of the distribution $\beta(\Psi)$, plays
an important role.  So, for different angular velocity distributions
in the open magnetic field line zone there should exist different
magnetosphere configurations that have in general distinct current
density distributions.  Let us now consider restrictions on the
differential rotation rate $\beta(\Psi)$.

%

\subsection{Angular velocity of the open magnetic field lines}
\label{sec:Beat(psi)}

Due to rotation of the NS a large potential difference arises between
magnetic field line foot points at the surface of the NS.  The
potential difference between the pole and the magnetic field line
corresponding to the value of the magnetic flux function $\Psi$ is
\begin{equation}
  \label{eq:Vpsi}
  \Delta\mathcal{V}(\Psi) = \frac{\mu}{\RLC^2}\Psi 
\end{equation}
In perfectly force-free magnetosphere the magnetic field lines are
equipotential.  However, due to presence of the polar cap acceleration
zone, where MHD conditions are broken a part of this potential
difference appears as a potential drop between the surface of the NS
and the pair-formation front, above which the magnetic field line
remains equipotential.  This potential drop is the reason why the open
magnetic field lines rotate differently from the NS.  The normalised
angular velocity of a magnetic field line $\beta$ is expressed trough
the potential drop along the field line as (e.g.  \citet{BeskinBook},
\PapI)
\begin{equation}
  \label{eq:OmF}
  \beta = 1 + \frac{\RLC^2}{\mu}\frac{\pd \mathpzc{V}(\Psi)}{\pd \Psi}
  \,.
\end{equation}
$\mathpzc{V}$ is the total potential drop (in statvolts) along the
magnetic field line in the polar cap acceleration zone (cf. eq.~(23)
in \PapI).

The polar cap of pulsar is limited by the magnetic field line
corresponding to a value of the flux function $\PsiL$.  The potential
drop between the rotation axis and the boundary of the polar cap is
\begin{equation}
  \label{eq:Vvac}
  \Delta\mathcal{V}(\PsiL) = \frac{\mu}{\RLC^2}\PsiL \equiv \Delta\PC{\mathcal{V}} 
\end{equation}
This is the maximum available potential drop along an open magnetic
field line. It could be achieved in vacuum, when there is no plasma in
the polar cap.  We will call $\Delta\PC{\mathcal{V}}$ the vacuum
potential drop.  Let us normalise the poloidal flux function $\Psi$ to
its value at the last open magnetic field line $\PsiL$ and introduce a
new function $\psi\equiv\Psi/\PsiL$.  Normalising potential drop along
field lines to the vacuum potential drop and introducing the
dimensionless function $V\equiv{}\mathpzc{V}/\Delta\PC{\mathcal{V}}$
we rewrite expression for the normalised angular velocity of of the
open magnetic field line as
\begin{equation}
  \label{eq:BetaV}
  \beta = 1 + \frac{\pd V}{\pd\psi}
  \,.  
\end{equation}
As the potential drop \emph{along} any field line can not be greater
than the vacuum drop and could not have different sign than the vacuum
drop, variation of the electric potential \emph{across} the polar cap
can not exceed the vacuum potential drop. In terms of the
dimensionless functions this condition has the form
\begin{equation}
  \label{eq:DeltaV_less_1}
  |V(\psi_1)-V(\psi_2)| \le 1, \quad \forall\,\psi_1,\psi_2 \in [0,1]  
  \,.  
\end{equation}
Inequality~(\ref{eq:DeltaV_less_1}) sets the limit on the electric
potential in the polar cap of pulsar.

\subsection{Current density in the polar cap}
\label{sec:S(psi)}

In order to obtain the current density distribution in the polar cap
of pulsar the pulsar equation~(\ref{eq:PsrEq}) together with the
condition at the light cylinder~(\ref{eq:CondAtLC}) must be solved.
There is an analytical solution of the pulsar equation only for
split-monopole configuration of the poloidal magnetic field.  Namely,
when flux function $\Psi$ has the form
\begin{equation}
  \label{eq:PsiMonopole}
  \Psi =\PsiM (1-\cos\theta)
  \,,
\end{equation}
$\PsiM$ being a constant, equations~(\ref{eq:PsrEq}) and
(\ref{eq:CondAtLC}) have a smooth solution if the poloidal current
function $S$ has the form \citep[e.g.][]{Blandford/Znajek77}
\begin{equation}
  \label{eq:SMonopole}
  S(\Psi) = - \beta(\Psi)\, \Psi (2-\frac{\Psi}{\PsiM})
  \,.
\end{equation}
Here $\PsiM$ corresponds to the value of the magnetic flux trough the
upper hemisphere, i.e. it corresponds to the magnetic field line lying
in the equatorial plane.  The poloidal current given by
equation~(\ref{eq:SMonopole}) is very similar to current in the well
known Michel solution \citep{Michel73}, but this expression is valid
for non-constant $\beta(\Psi)$ too.

In this paper we will use for the poloidal current function $S$
expression~(\ref{eq:SMonopole}).  Doing so, we assume that in the
neighbourhood of the light cylinder the geometry of the poloidal
magnetic field is close to a split monopole.  This is good
approximation if the size of the closed magnetic field line zone is
much smaller than the light cylinder size, $x_0\ll1/\beta(\psi),\
\psi<1$.  For configurations where the size of the corotating zone%
\footnote{plasma in the closed field line zone corotates with the NS,
  so we will call the region with the closed magnetic field lines the
  corotating zone}
approaches the light cylinder the poloidal current $S$ is different
from the one given by eq.~(\ref{eq:SMonopole}), but we expect that
this deviation should not exceed 10-20 per cents.  Indeed, in the
numerical simulations described in \PapI{}, where the case of constant
$\beta\equiv1$ was considered, the deviation of $S$ from the Michel's
poloidal current did not exceed 20 per cents and it got smaller for
smaller size of the corotating zone (see Fig.~3 in \PapI{}).
Similarly, in \PapII{}, where we considered the case of variable
$\beta<1$, the poloidal current deviated from the values given by the
analytical formula~(\ref{eq:SMonopole}) by less than 20 per cents and
the difference became smaller for smaller size of the corotating zone.
So, we may hope that the same relation holds in the general case too.

We intent to find the range of admitted current density distributions
in the force-free magnetosphere.  Here we use the split-monopole
approximation for the poloidal current~(\ref{eq:SMonopole}), hence, we
can study only the effect of differential rotation on the current
density distribution.  The dependence of the current density on the
size of the corotating zone in differentially rotating magnetosphere
will be addressed in a subsequent paper, where we will refine our
results by performing numerical simulations for different sizes of the
corotating zone.

So, in our approximation the last closed field line in dipole geometry
corresponds to the field line lying in the equatorial plane in
monopole geometry, i.e. $\PsiM=\PsiL$.  In normalised variables
expression for the poloidal current has the form
\begin{equation}
  \label{eq:SMonopoleNormalized}
  S(\Psi) = - \PsiL\, \beta(\psi)\, \psi (2-\psi)
  \,.
\end{equation}
The poloidal current density in the magnetosphere is \citep[see
e.g.][]{BeskinBook}
\begin{equation}
  \label{eq:j_general}
  \Pol{j} = \frac{c}{4\pi}\frac{\mu}{\RLC^4}
  \frac{\vec{\nabla} S \times \vec{e_\phi}}{r\sin\theta}=
  \frac{\Omega \Pol{\vec{B}}}{2\pi{}c} c \: \frac{1}{2}\frac{d S}{d\Psi}
  \,.
\end{equation}
In the polar cap of pulsar the magnetic field is dipolar and, hence
poloidal. The Goldreich-Julian charge density for the corotating
magnetosphere near the NS is
\begin{equation}
  \label{eq:RhoGJ}
  \GJNS{\rho} = - \frac{\vec{\Omega}\cdot\vec{B}}{2\pi{}c}
  \,.
\end{equation}
Using  expressions~(\ref{eq:SMonopoleNormalized})-(\ref{eq:RhoGJ})
we get for the current density in the polar cap of pulsar
\begin{equation}
  \label{eq:j}
  j =  \frac{1}{2} \GJNS{j}
  \left[
    2\beta(1-\psi) + \beta'\psi(2-\psi)
  \right]
  \,.
\end{equation}
The prime denotes differentiation with respect to $\psi$, i.e.
$\beta'\equiv d\beta/d\psi$; $\GJNS{j}\equiv\GJNS{\rho}\,c$ is the
Goldreich-Julian current density in the polar cap for the
\emph{corotating} magnetosphere.  At the surface of the NS, where the
potential drop is zero and plasma corotates with the NS, $\GJNS{j}$
corresponds to the local GJ current density.

\subsection{Goldreich-Julian charge density in the polar cap for
  differentially rotating magnetosphere}
\label{sec:RhoGJ_local}

The Goldreich-Julian (GJ) charge density is the charge density which
supports the force-free electric field:
\begin{equation}
  \label{eq:RhoGJ_local_E}
  \GJ{\rho} \equiv  \frac{1}{4\pi}\, \vec{\nabla}\cdot\vec{E}
  \,.
\end{equation}
The GJ charge density in points along a magnetic field line rotating
with an angular velocity different from the angular velocity of the NS
will be different from values given by the eq.~(\ref{eq:RhoGJ}).
Substituting expression for the force-free magnetic field~(\ref{eq:E})
into eq.~(\ref{eq:RhoGJ_local_E}) we get
\begin{equation}
  \label{eq:RhoGJ_local_Psi}
  \GJ{\rho}  =  - \frac{\mu}{4\pi\RLC^4}\,
  (\beta \Delta\Psi + \beta'(\nabla\Psi)^2)
  \,.
\end{equation}
We see that the GJ charge density depends not only on the angular
velocity of the field line rotation (the first term in
eq.~(\ref{eq:RhoGJ_local_Psi})), but also on the angular velocity
profile (the second term in eq.~(\ref{eq:RhoGJ_local_Psi})).

Near the NS the magnetic field is essentially dipolar.  The magnetic
flux function $\Psi$ for dipolar magnetic field is
\begin{equation}
  \label{eq:Psi_Dipole}
  \Psi^\mathrm{dip} = \frac{\sin^2\theta}{r}
  \,.
\end{equation}
Substituting this expression into equation~(\ref{eq:RhoGJ_local_Psi})
we get
\begin{eqnarray}
  \label{eq:RhoGJ_local_theta}
  \GJ{\rho} & = &  
  - \frac{\mu}{4\pi\RLC^4}\,\frac{1}{r^3} \times 
  \nonumber \\
  &&
  \left(
    \beta\, 2 (3\cos^2\theta-1) + 
    \beta' \frac{\sin^2\theta}{r} (3\cos^2\theta+1)
  \right)
  \,.
\end{eqnarray}
In the polar cap of pulsar $\cos\theta\simeq{}1$ and
$\mu/(r\RLC)^3\simeq{}B/2$.  Recalling expression for the magnetic
flux function for dipole magnetic field~(\ref{eq:Psi_Dipole}) we get
for the local GJ charge density in the polar cap of pulsar
\begin{equation}
  \label{eq:RhoGJ_local}
  \GJ{\rho} = \GJNS{\rho}\, (\beta + \beta'\psi)
  \,.
\end{equation}

\section{Accelerating potential}
\label{sec:Equation_for_V}

In our approximation any current density distribution in force-free
magnetosphere of an aligned rotator has the form given by
eq.~(\ref{eq:j}).  The current density depends on the angular velocity
of the magnetic field lines $\beta(\psi)$, which for a given field
line depends on the total potential drop along that line via
equation~(\ref{eq:BetaV}).  The potential drop in the acceleration
zone can not exceed the vacuum potential drop, i.e. $V$ is limited by
inequality~(\ref{eq:DeltaV_less_1}).

So, if we wish to find the accelerating potential which supports a
force-free configuration of the magnetosphere for a given form of the
current density distribution%
\footnote{guessed from a model for the polar cap cascades, for
  example}
in the polar cap we do the following.  We equate the expression for
the current density distribution to the general expression for the
current density~(\ref{eq:j}), then we express $\beta(\psi)$ in terms
of $V(\psi)$ by means of equation~(\ref{eq:BetaV}), and obtain thus an
equation for the electric potential $V$ which supports a force-free
magnetosphere configuration with the desired current density
distribution.  If solutions of the obtained equation fulfil
limitation~(\ref{eq:DeltaV_less_1}), such configuration is admitted,
if not -- such current density could not flow in force-free
magnetosphere of an aligned pulsar.  Currently there is no detailed
model for non-stationary polar cap cascades from which we could deduce
reasonable shapes for current density distribution. Therefore, we try
to set constrains on the current density assuming linear dependence of
the current density on $\psi$.

In differentially rotating magnetosphere there are two characteristic
current densities.  The first one is the Goldreich-Julian current
density for the corotating magnetosphere $\GJ{j}^0$.  It corresponds to
the actual Goldreich-Julian current density in the magnetosphere at
the NS surface, where differential rotation is not yet build up.  The
second characteristic current density is the actual Goldreich-Julian
current density $\GJ{j}$ in points above the acceleration zone where
the magnetosphere is already force-free and the final form of
differential rotation is established; in the polar cap $\GJ{j}$ is
given by formula~(\ref{eq:RhoGJ_local}).  For magnetosphere with
strong differential rotation the current densities $\GJ{j}^0$ and
$\GJ{j}$ differ significantly.  In this section we consider both
cases, namely, when the current density distribution is normalised to
$\GJ{j}^0$ and when it is normalised to $\GJ{j}$.

\subsection{Outflow with the current density being a constant fraction
  of the actual Goldreich-Julian current density}

For non-stationary cascades the physics would be determined by the
response of the cascade zone to the inflowing particles and MHD waves
coming from the magnetosphere.  However, the accelerating electric
field depends on the deviation of the charge density from the local
value of the GJ charge density.  So, the first naive guess would be
that the preferable state of the cascade zone would be the state when
(on average) the current density is equal to the GJ current density
$\GJ{j}$:
\begin{equation}
  \label{eq:j=jGJ}
  j(\psi) = \GJ{j}(\psi) = \GJNS{j}\, (\beta + \beta'\psi)
  \,.
\end{equation}
Equating this formula to the general expression for the current
density~(\ref{eq:j}) and substituting for $\beta$
expression~(\ref{eq:BetaV}), after algebraical transformation we get
the equation for the accelerating electric potential in the polar cap
of pulsar
\begin{equation}
  \label{eq:Eq_V_for_jGJ_local}
  V'' = - 2\, \frac{1+V'}{\psi}
  \,.
\end{equation}
We set the boundary conditions for $V$ at the edge of the polar cap.
As the boundary conditions we can use the value of the normalised
angular velocity at the edge of the polar cap and the value of the
electric potential there
\begin{eqnarray}
  \label{eq:Eq_V_BoundaryCond_beta}
  1+V'(1) &=& \PC{\beta} \\ 
  \label{eq:Eq_V_BoundaryCond_V}
  V(1) &=& V_0
\end{eqnarray}
Solution of equation~(\ref{eq:Eq_V_for_jGJ_local}) satisfying the
boundary
conditions~(\ref{eq:Eq_V_BoundaryCond_V}),~(\ref{eq:Eq_V_BoundaryCond_beta})
is
\begin{equation}
  \label{eq:SolV_for_jGJ_local}
  V(\psi) = V_0 + (1-\psi) \left( 1 - \frac{\PC{\beta}}{\psi} \right)
  \,.
\end{equation}
We see that unless $\PC{\beta}=0$ the potential has singularity at the
rotation axis, and, hence, such configuration can not be realised in
force-free magnetosphere of a pulsar.  The
condition~(\ref{eq:DeltaV_less_1}) is violated -- the potential
difference exceeds the vacuum potential drop.

If $\PC{\beta}=0$, the potential is $V=V_0 + 1-\psi$ and from
eq.~(\ref{eq:BetaV}) we have $\beta(\psi)\equiv{}0$.  Substituting
this into eq.~(\ref{eq:j}) we get for the current density
$j(\psi)\equiv{}0$.  So, the case with $\PC{\beta}=0$ is degenerated,
as there is no poloidal current in the magnetosphere, it corresponds
to the vacuum solution.

Let us consider now a more general form for the current density
distribution
\begin{equation}
  \label{eq:j=A_jGJ}
  j(\psi) = A \GJ{j}(\psi) = A \GJNS{j}\, (\beta + \beta'\psi)
  \,,
\end{equation}
where $A$ is a constant. In that case for the accelerating electric
potential in the polar cap of pulsar we have the equation
\begin{equation}
  \label{eq:EqV_for_AjGJ_local}
  V'' = 2(1+V') \frac{1-A-\psi}{\psi\left[\psi+2(A-1)\right]}
\end{equation}
For the same boundary
conditions~(\ref{eq:Eq_V_BoundaryCond_V}),~(\ref{eq:Eq_V_BoundaryCond_beta})
solution of this equation is
\begin{equation}
  \label{eq:SolGen_V_jGJ_local}
  V(\psi) = V_0 + 1 - \psi  + 
  \frac{\PC{\beta}( 2 A - 1 )}{2(A -1)}
  \log\left[
      \frac{\psi (2 A - 1)}{\psi + 2 (A - 1)}
  \right]
  .
\end{equation}
This solution is valid for $A\neq{}1,1/2$.  There is the same problem
with the electric potential in that solution.  Namely, unless
$\PC{\beta}=0$ the potential $V$ is singular%
\footnote{the singularity arises because $V''(0)$ goes to infinity
  unless $1+V'(0)=\beta(0)$ is zero, as it follows from
  equation~(\ref{eq:EqV_for_AjGJ_local})}
at the rotation axis.  The case with $A=1/2$ is also degenerated,
because in that case the solution for the electric potential is
$V(\psi)=V_0+1-\psi$ what yield the current density
$j(\psi)\equiv{}0$.

We see that solutions with the current density being a constant
fraction of the actual GJ current density are not allowed, except a
trivial degenerated case, corresponding to no net particle flow.  The
naive physical picture does not work and the current density in the
magnetosphere in terms of the actual GJ current density must vary
across the polar cap.  On the other hand, the GJ current density is
itself a variable function across the polar cap, it changes also with
the altitude within the acceleration zone, when the potential drop
increases until it reaches its final value.  So, we find it more
convenient to consider the current density in terms of the
corotational GJ current density.

\subsection{Outflow with the current density being a linear function
  of the magnetic flux in terms of the corotational Goldreich-Julian
  current density}

In models with the space charge limited flow (SCLF), when charged
particles can freely escape from the NS surface
\citep[e.g.][]{Arons/Scharlemann/78}, the charge density at the NS
surface is always equal to the local GJ charge density there,
$(\rho=\GJNS{\rho})|_{r=\RNS}$.  For SCLF the actual current density
in the polar cap could be less than $\GJNS{j}$ if acceleration of the
particles is periodically blocked in the non-stationary cascades.  The
current density could be greater than $\GJNS{j}$ if there is an inflow
of particles having opposite charge to that of the GJ charge density
from the magnetosphere into the cascade zone \citep[e.g.][]{Lyubar92}.
Therefore, an expression for the current density in terms of the
corotational GJ current density $\GJNS{j}$ would be more informative
from the point of view of the cascade physics.

Let us consider the case when the current density distribution in the
polar cap of pulsar has the form
\begin{equation}
  \label{eq:j=a_psi+b}
  j = \GJNS{j}(a\psi+b) 
  \,,
\end{equation}
where $a,b$ are constants.  The Michel current density distribution is
a particular case of this formula and corresponds to the values of
parameters $a=-1,b=1$.  The equation for the electric potential for
this current density is
\begin{equation}
  \label{eq:EqV_for_jGJ_NS}
  V'' = 2\,\frac{a\psi+b-(1+V')(1-\psi)}{\psi(2-\psi)}
  \,.
\end{equation}
Solution of the equation~(\ref{eq:EqV_for_jGJ_NS}) satisfying  the boundary
conditions~(\ref{eq:Eq_V_BoundaryCond_V}),~(\ref{eq:Eq_V_BoundaryCond_beta})
is
\begin{multline}
  \label{eq:SolGen_V_jGJ_NS}
  V(\psi) =  
  V0 + (1+a)(1-\psi) + \\
  + \frac{1}{2} 
  \log\left[
    (2-\psi )^{-\PC{\beta}-3a-2b} 
    \psi ^{\PC{\beta}-a-2b}
  \right]
  .
\end{multline}
We see that the potential is non singular at the rotation axis, if
$\PC{\beta}=a+2b$.  So, the admitted solution for the electric
potential is
\begin{equation}
  \label{eq:Sol_V_jGJ_NS}
  V(\psi) = V0 + (1+a)(1-\psi) - 2(a+b)\log(2-\psi )
  \,.
\end{equation}
In the rest of the paper we will use for the electric potential
expression~(\ref{eq:Sol_V_jGJ_NS}).  We will analyse physical
properties of force-free magnetosphere configurations when the
electric potential in the acceleration zone of the polar cap has that
form.

\section{Properties of admitted configurations}
\label{sec:main-results}

\subsection{Admitted current density}
\label{sec:admitted-current}

The potential drop in the polar cap of pulsar is limited by the vacuum
potential drop.  In our notations this limit is formulated as
inequality~(\ref{eq:DeltaV_less_1}).  Parameters $a,b$ from the
expression for the electric current~(\ref{eq:j=a_psi+b}) enters into
the formula for the electric potential~(\ref{eq:Sol_V_jGJ_NS}).
Imposing limitation~(\ref{eq:DeltaV_less_1}) we get the admitted range
for these parameters in the force-free magnetosphere.  In
Appendix~\ref{sec:App--admitt-curr-dens} we do such analysis and find
the region in the plane $(a,b)$ which is admitted by the
requirement~(\ref{eq:DeltaV_less_1}).  This region is shown as a grey
area in Fig.~\ref{fig:ab_range_full}.  From
Fig.~\ref{fig:ab_range_full} it is evident that for the most of the
admitted values of parameters $a,b$ the current density has different
signs in different parts of the polar cap.  There is also a region
where the values of parameters correspond to the current density
distributions having the same sign as the GJ charge density in the
whole polar cap.

The physics of the polar cap cascades impose additional limitations on
the current density and accelerating electric potential distribution
in the polar cap.  There is now no detailed theory of non-stationary
polar cap cascades.  Therefore, in setting constrains on the current
density distribution we should use some simple assumptions about the
possible current density.  There is a preferable direction for the
accelerating electric field in the polar cap.  The direction of this
field in such that it accelerates charged particles having the same
sign as the GJ charge density away from the star.  It is natural to
assume that the average current in the polar cap cascade should flow
in the same direction.  The average current could flow in the opposite
direction only if the accelerating electric field is screened.  In
order to screen the accelerating field a sufficient amount particles
of the same sign as the accelerated ones should come from the
magnetosphere and penetrate the accelerating potential drop.  These
particles, however, are itself produced in the polar cap cascade.
They must be accelerated somewhere in the magnetosphere back to the NS
up to the energy comparable with the energy the primary particles get
in the polar cap cascade.  Even if the problem of particle
acceleration back to the NS could be solved, screening of the electric
field will interrupt the particle creation, and, hence, there will be
not enough particles in the magnetosphere which could screen the
electric filed the next time.  Although the real physics is more
complicated and is not yet fully understood, the case of the
unidirectional current in the polar cap is worth of detailed
investigation as it is ``the most natural'' from the point of view of
the polar cap cascade physics.  In the following we will call the
current being of the same sign as the GJ charge density as
``positive'' and the current being of the opposite sign to the GJ
charge density as ``negative''.

\begin{figure}
  \includegraphics[clip,width=\columnwidth]{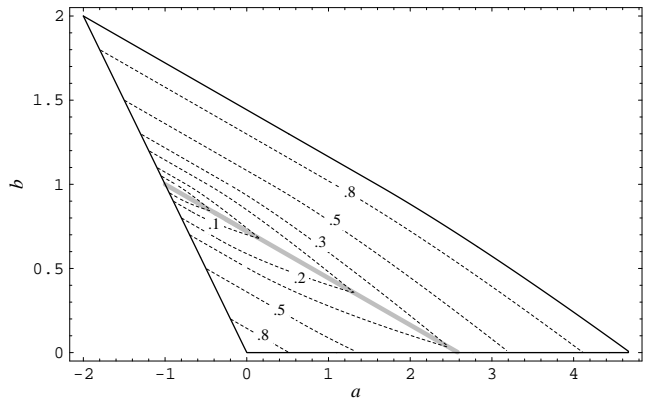}
  \caption{Maximum potential drop across the polar cap. The dotted
    lines show contours of $\Max{\Delta{}V}$.  Contours for
    $\Max{\Delta{}V}=0.05,0.1,0.2,0.5,0.8$ are shown.  Labels on the
    lines correspond to the values of $\Max{\Delta{}V}$.  The line
    corresponding to $\Max{\Delta{}V}=0.05$ is not labelled.}
  \label{fig:Vmax_Contours}
\end{figure}

The linear current density distribution~(\ref{eq:j=a_psi+b}) will be
always positive if
\begin{equation}
  \label{eq:ab_for_j_positive}
  b\ge\max(-a,0)
  \,.
\end{equation}
Only a subset of the admitted values of $a,b$ corresponds to the
positive current density distribution.  Such values of the parameters
$a,b$ are inside the triangle-like region shown in
Figs.~\ref{fig:Vmax_Contours},~\ref{fig:j/jGJ_Min},~\ref{fig:W}.  We
see that a rather wide variety of positive current density
distributions are admitted in the force-free magnetosphere: current
density distributions being constant across the polar cap of pulsar
are admitted, as well as current densities decreasing or increasing
toward the polar cap boundary.  So, the current density in the
force-free magnetosphere could deviate strongly from the classical
Michel current density, corresponding to the point $a=-1, b=1$.  The
price for this freedom is the presence of a non-zero accelerating
electric potential in the polar cap.  If the price for a particular
current density distribution is too hight, i.e. if the potential drop
is too large, only magnetosphere of pulsars close to the ``death
line'' could admit such current density.  Let us now consider the
distribution of the potential drop in the parameter space $(a,b)$.

%

\subsection{Electric potential}
\label{sec:electric-potential}

We emphasised already that the shape of the function $V(\psi)$ is very
important for the resulting current density distribution.  However, as
we do not understand in detail the physics of non-stationary cascades,
we cannot judge whether a particular form of $V(\psi)$ is admitted by
the cascade physics or not.  On the other hand, in young pulsars the
average potential drop could not be very large, because already a
small fraction of the vacuum potential drop would be sufficient for
massive pair creation and screening of the accelerating electric
field.  So, currently we could judge about reasonableness of a
particular current density distribution only from the maximum value of
the potential drop it requires.  The electric potential given by
eq.~(\ref{eq:Sol_V_jGJ_NS}) is known up to the additive constant
$V_0$, which is the value of the accelerating potential at the polar
cap boundary.  $V_0$ and thus the actually potential drop in the
accelerating zone can not be inferred from the magnetosphere physics
and is set by the physics of the polar cap cascades.  The only thing
we could say about the actual potential drop in the acceleration zone
\emph{along} field lines is that its absolute value is not smaller
than the absolute value of the maximum potential drop of $V(\psi)$
\emph{across} the polar cap.

Let us now consider possible values of the maximum potential drop
across the polar cap of pulsar.  If the potential is a monotone
function of $\psi$ in the polar cap, the maximum potential drop is the
drop between the rotation axis and the polar cap boundary.  If the
potential as a function of $\psi$ has a minimum inside the polar cap,
the maximum potential drop will be either between the axis and the
minimum point, or between the edge and the minimum point.  We analyse
this issue in details in Appendix~\ref{sec:App-maximum-potent-drop}.
In Fig.~\ref{fig:Vmax_Contours} the contour map of the maximum
potential drop in the plane $(a,b)$ is shown.  The line given by
eq.~(\ref{eq:DVb_is0}) is the line where for fixed $a$ (or $b$) the
smallest value of the potential drop across the polar cap is achieved.
From this plot it is evident that even if the potential drop in the
polar cap is rather moderate, of the order of $\sim{}10$ per cents,
there are force-free magnetosphere configurations with the current
density distribution deviating significantly from the Michel current
density distribution.  So, even for young pulsars there may be some
flexibility in the current density distribution admitted by the
force-free electrodynamics.

Note that force-free magnetosphere impose different constraints on
pulsars in aligned $\vec{\mu}\cdot\vec{\Omega}>0$ and anti-aligned
configuration $\vec{\mu}\cdot\vec{\Omega}<0$ configuration (pulsar and
antipulsar in terms of \citet{Ruderman/Sutherland75}).  For pulsars
the accelerating potential is positive, i.e. it increases from the
surface of the NS toward the force-free zone above the pair formation
front.  In case of antipulsar the potential is negative, it decreases
toward the pair formation front, because positive charges are
accelerated.  Equations for the current
density~(\ref{eq:j}),~(\ref{eq:j=a_psi+b}) we used to derive the
equation for the electric potential~(\ref{eq:EqV_for_jGJ_NS}) contain
the expression for the GJ charge density as a factor, and, hence, the
resulting expression for the electric potential is the same for each
sign of the GJ current density.  So, for pulsars there is a
\emph{minimum} in the accelerating potential distribution, for
antipulsar the distribution of the accelerating electric potential has
a \emph{maximum}.  Mathematically this results from different signs of
the integration constant $V_0$.

\subsection{Angular velocity}
\label{sec:angular-velocity}

%
\begin{figure}
  \includegraphics[clip,width=\columnwidth]{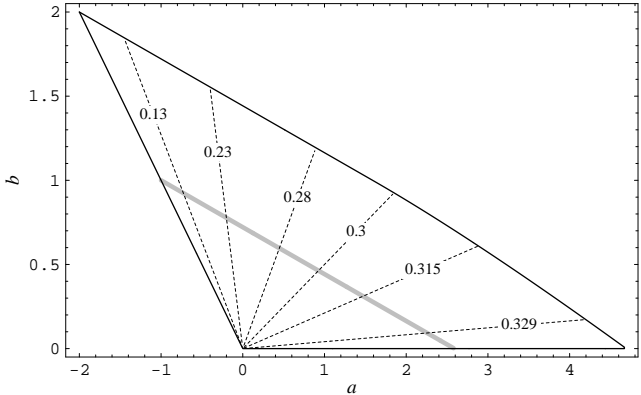}
  \caption{Ratio of the actual current density to the Goldreich-Julian
    current density $\iota(1)$ at the polar cap boundary, where the
    minimum value of this ratio is achieved, see text.  The dotted
    lines show contours of $\iota(1)$}
  \label{fig:j/jGJ_Min}
\end{figure}

The normalised angular velocity of the open magnetic field lines in
the force free magnetosphere with linear current density
distribution~(\ref{eq:j=a_psi+b}) is given by
\begin{equation}
  \label{eq:Beta}
  \beta(\psi) =\frac{2 b + a \psi }{2-\psi }
  \,.
\end{equation}
For admitted current densities it grows with increasing of $\psi$,
because the first derivative $d\beta/d\psi$ for the admitted values of
$a,b$ is always non-negative.  So, the angular velocity either
\emph{increase} toward the polar cap boundary or remains
\emph{constant} over the cap if $a=-b$. The latter case includes also
the Michel solution.  The minimum value of $\beta$
\begin{equation}
  \label{eq:Beta_Min}
  \beta_\mathrm{min} = b
  \,,
\end{equation}
is achieved at the rotation axis, where $\psi=0$, and the maximum value
\begin{equation}
  \label{eq:Beta_Max}
  \beta_\mathrm{min} = 2 b + a
  \,,
\end{equation}
at the polar cap boundary, where $\psi=1$.  So, the open field lines
can rotate slower, as well as faster that the NS, but the lines near
the polar cap boundary rotate not slower than the lines near the
rotation axis.

\subsection{Goldreich-Julian current density}
\label{sec:goldr-juli-curr}

%
\begin{figure}
  \includegraphics[clip,width=\columnwidth]{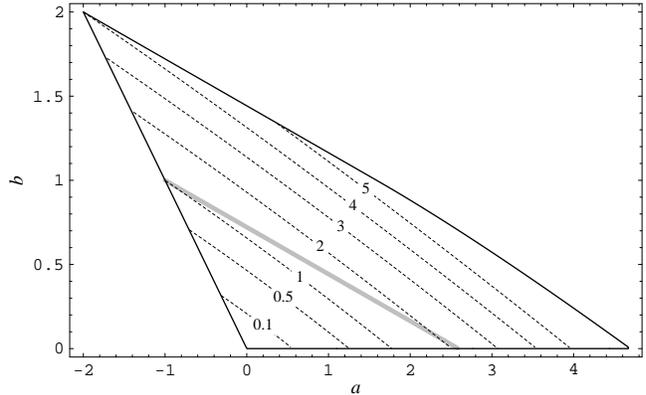}
  \caption{Spindown rate in terms of the Michel spindown rate. Dotted
    lines show contours of $w$.  Label on the lines correspond to the
    values of $w$.}
  \label{fig:W}
\end{figure}

Expression for the GJ current density in the polar cap can be obtained
by substitution of the expression~(\ref{eq:Beta}) for $\beta$ into
equation~(\ref{eq:RhoGJ_local}) for the GJ current density. We get
\begin{equation}
  \label{eq:RhoGJ_linear_j}
  \GJ{j}(\psi) = \GJNS{j}\, \frac{4b + a\psi(4-\psi)}{(\psi -2)^2}
  \,.
\end{equation}
For the admitted values of the parameters $a,b$ the derivative
$d\GJ{j}/d\psi$ is always non-negative and, hence, the GJ current
density either \emph{increases} toward the polar cap boundary, or
remains \emph{constant} when $a=-b$.  The actual current density,
however, could decrease as well as increase toward the polar cap edge.

For the charge separated flow the deviation of the current density
from the GJ current density generate an accelerating or a decelerating
electric field when $j<\GJ{j}$ or $j>\GJ{j}$ correspondingly.
Although in non-stationary cascades the particle flow would be not
charge separated, the ratio of the actual current density to the GJ
current density may give some clues about cascade states required by a
particular magnetosphere configuration.  This ratio is given by
\begin{equation}
  \label{eq:j_to_jGJ}
  \iota(\psi) \equiv \frac{j(\psi)}{\GJ{j}(\psi)} = 
  \frac{(\psi -2)^2 (b+a \psi )}{a \psi (4-\psi)  +4 b}
\end{equation}
For each admitted configuration the current density is equal to the GJ
current density at the rotation axis.  For the admitted values of the
parameters $a,b$ the derivative $d\,\iota/d\psi$ is always positive,
and, hence, the current density in terms of the GJ current density
\emph{decreases} toward the polar cap boundary.  So, except the
rotation axis the current density in the polar cap is always less than
the GJ current density.  The relative deviation of the actual current
density from the GJ current density is maximal at the polar cap
boundary
\begin{equation}
  \label{eq::j_to_jGJ_min}
  \iota(1) = \frac{a+b}{3 a+4 b}
  \,.
\end{equation}
Its maximum value $\iota_\mathrm{max}=1/3$ this ratio achieves when
$b=0$.  Its minimum value $\iota_\mathrm{min}=0$ it achieves when
$a=-b$, what includes also the case of the Michel's current density
distribution.  The contours of $\iota(1)$ are shown in
Fig.~\ref{fig:j/jGJ_Min}.

\subsection{Spin-down rate and the total energy of electromagnetic
  field in the magnetosphere}

%
\begin{figure}
  \includegraphics[clip,width=\columnwidth]{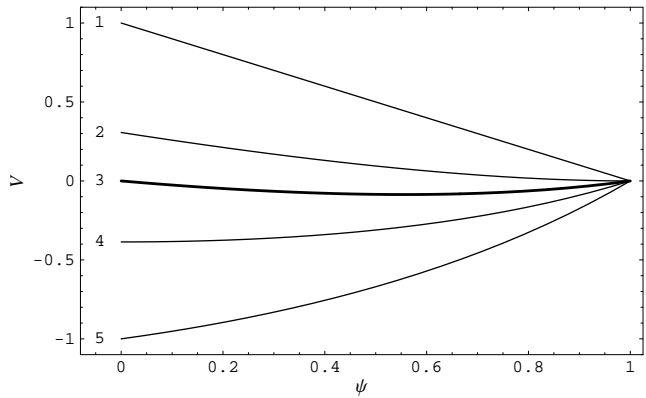}
  \caption{Electric potential in the polar cap of pulsar as a function
    of the normalised flux function $\psi$ for magnetosphere
    configurations with a constant current density across the cap.  In
    all cases $V_0$ is set to zero.  Numbers near the lines correspond
    to the following values of $b$: 1 --- $b=0$; 2 --- $b=.5$; 3 ---
    $b=\Max{b}/2$; 4 --- $b=1$; 5 --- $b=\Max{b}$.  The line
    corresponding to the minimum potential drop across the cap is
    shown by the thick solid line (the line 3).}
  \label{fig:a=0_v}
\end{figure}

In our notations the spindown rate of an aligned rotator is
(cf. eq.~(60) in \PapI)
\begin{equation}
  \label{eq:W_general}
  W = \Wmd \int_0^{\PsiL} S(\Psi) \beta(\Psi)\, d\Psi
  \,,
\end{equation}
where $\Wmd$ is the magnetodipolar energy losses defined as
\begin{equation}
  \label{eq:Wmd}
  \Wmd = \frac{B_0^2\RNS^6\Omega^4}{4c^3}
  \,.
\end{equation}
Substituting expression for the poloidal
current~(\ref{eq:SMonopoleNormalized}) and using the normalised flux
function $\psi$ we get
\begin{equation}
  \label{eq:W}
  W = \Wmd\, \PsiL^2 \int_0^1 \beta^2(\psi) \psi(2-\psi)\, d\psi
  \,.
\end{equation}

Expression for the spindown rate in the Michel solution
\begin{equation}
  \label{eq:W_Michel}
  W_\mathrm{M} = \frac{2}{3} \PsiL^2 \Wmd
\end{equation}
differs from the spindown rate obtained in the numerical simulations
of the corotating aligned rotator magnetosphere with by a constant factor.
However, it has very similar dependence on the size of the corotating
zone $x_0$ (cf. equations~(62),~(63) in \PapI).  As our solutions are
obtained in split-monopole approximation, they should differ from the
real solution approximately in the same way as the Michel solution
does.  Because of this it would be more appropriate to normalise the
spindown rate to the spindown rate in the Michel split-monopole
solution.  Doing so we will be able to study the effect of
differential rotation on the energy losses separately from the
dependence of the spindown rate on the size of the corotating zone.

For the normalised spindown rate in the considered case of linear
current density we get
\begin{multline}
  \label{eq:W_to_W_Michel}
  w \equiv \frac{W}{W_\mathrm{M}} =
  4 a^2 (3 \log2 - 2) + \\
  + 3 a b (8 \log2-5) + 6 b^2 (2 \log2-1) 
  \,.
\end{multline}
In Fig.~\ref{fig:W} contour lines of $w$ are shown in the domain of
admitted values for parameters $a,b$.  We see that spindown rate can
vary significantly, from zero to the value exceeding the Michel's
energy losses by a factor of $\approx6$.  It increases with increasing
of the values of the parameters $a,b$ and decreases with decreasing of
that values.  It is due to increasing or decreasing of the total
poloidal current in the magnetosphere correspondingly.

The total energy of the magnetosphere could be estimated from the
split-monopole solution.  Using the formula~(\ref{eq:W__Spindown})
derived in Appendix~\ref{sec:energy-electr-field} we have for the
total energy of the electromagnetic field
\begin{equation}
  \label{eq:W_total}
  \mathcal{W}\simeq\Pol{\mathcal{W}} + (R-\RNS)\,W
  \,,
\end{equation}
where $\Pol{\mathcal{W}}$ is the total energy of the poloidal magnetic
field and $R$ is the radius of the magnetosphere.  The first term in
our approximation is the same for all magnetosphere configurations,
the difference in the total energy arises from the second term.
Hence, the contours of the constant total energy in the plane $(a,b)$
have the same form as the contours of the spindown rate $W$ shown in
Fig.~\ref{fig:W}.  So, the total energy of the magnetosphere increases
with increasing of parameters $a,b$, i.e. it increases with the
increase of the poloidal current.

%
\begin{figure}
  \includegraphics[clip,width=\columnwidth]{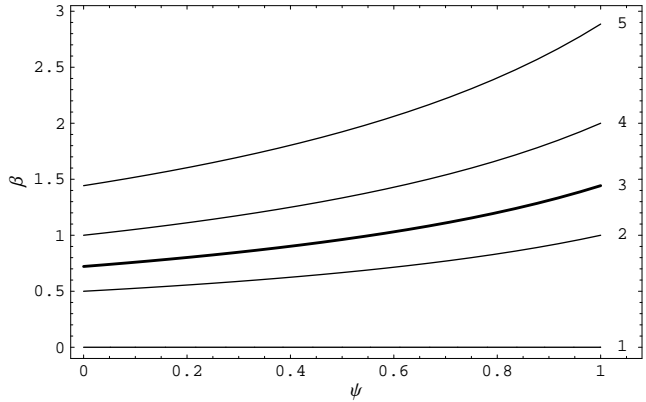}
  \caption{Normalised angular velocity of the open magnetic field
    lines as a function of the normalised flux function $\psi$ for
    magnetosphere configurations with a constant current density
    across the cap. Labelling of the curves is the same as in
    Fig.~\ref{fig:a=0_v}.}
  \label{fig:a=0_beta}
\end{figure}

\subsection{Example solutions}
\label{sec:part-solutions}

As examples we consider here the properties of two particular
solutions in details.  We chose these solution because either their
current density or the potential drop seem to correspond to
``natural'' states of the polar cap cascades.  Although we do not
claim that one of the solutions is realised as a real pulsar
configuration, but knowledge of their properties may be helpful in
understanding of the physical conditions the polar cap cascades should
adjust to.

\subsubsection{Configurations with constant current density}
\label{sec:constant-j}

%
\begin{figure}
  \includegraphics[clip,width=\columnwidth]{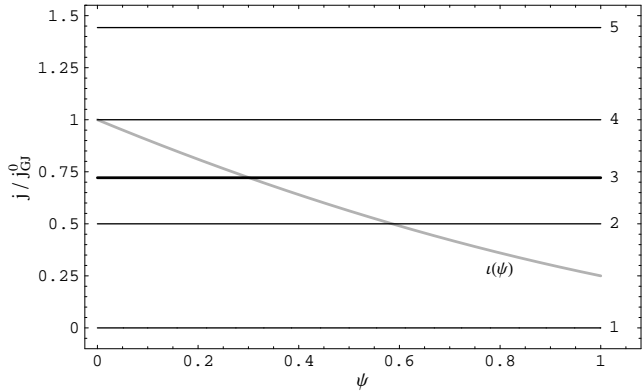}
  \caption{Current density as a function of the normalised flux
    function $\psi$ for magnetosphere configurations with a constant
    current density across the cap. Labelling of the curves is the
    same as in Fig.~\ref{fig:a=0_v}. By the thick grey line the ratio
    of the actual current density to the GJ current density
    $\iota(\psi)$ is shown. For this case it is the same for all
    solutions.}
  \label{fig:a=0_j}
\end{figure}

At first we consider the case when the current density is constant
over the polar cap, i.e. $a=0$ and $j=b\,\GJNS{j}$.  Constant density
distribution would be produced by cascades in their ``natural'' state,
if the current adjustment proceeds locally, without strong influence
from the current along adjacent field lines.  The electric potential
in that case is
\begin{equation}
  \label{eq:V_a_eq_0}
  \Const{V}(\psi) = V_0+1-\psi-2b\log(2-\psi)
  \,.
\end{equation}
This potential has the following properties (see Fig.~\ref{fig:a=0_v}
where $V(\psi)$ is shown for several values of $b$ assuming for the
sake of simplicity $V_0=0$):
\begin{itemize}
\item the admitted values of the current density in the polar cap of
  pulsar are within interval $[0,\Max{b}]$, where
  $\Max{b}=1/\log2\simeq{}1.443$.

\item if $0<b<\Max{b}/2\simeq{}0.721$ the value of the electric
  potential at the rotation axis $\Const{V}(0)$ is larger than that
  value at the polar cap edge $\Const{V}(1)$,
  $\Const{V}(0)>\Const{V}(1)$

\item if $\Max{b}/2<b<\Max{b}$ the value of the electric potential at
  the rotation axis $\Const{V}(0)$ is smaller than that value at the
  polar cap edge $\Const{V}(1)$, $\Const{V}(0)<\Const{V}(1)$.

\item if $0<b<1/2$ or $1<b<\Max{b}$ the potential is a monotone
  function of $\psi$; if $1/2<b<\Max{b}$ it has a minimum.

\item in the point $b=\Max{b}/2$ the maximum potential drop across the
  polar cap reaches its minimum value $\Max{\Delta{}V}=0.086$.
\end{itemize}
The reason for such behaviour of the potential is easy to understand
from the Fig.~\ref{fig:ab_range} in
Appendix~\ref{sec:App-maximum-potent-drop}.  The critical points where
$V(\psi)$ changes its behaviour are the points where the line $a=0$
intersects the boundaries of the regions I,II,II, and IV.

The angular velocity of the open magnetic field lines is
\begin{equation}
  \label{eq:Beta_b_eq_0}
  \Const{\beta}(\psi) = \frac{2 b}{2-\psi}
\end{equation}
Distribution of the corresponding angular velocity is shown in
Fig.~\ref{fig:a=0_beta}.  For $b>1$ the angular velocity of rotation
of all open magnetic field lines is larger than the angular velocity
of the NS. For $b<1/2$ all magnetic field lines rotate slower that the
NS. For $1/2<b<1$ some open filed lines near the rotation axis rotates
slower that the NS, the other open field lines rotates faster than the
NS.

The current density distribution in terms of the GJ current density is
\begin{equation}
  \label{eq:j2jGJ_a_eq_0}
  \Const{\iota}(\psi) = \frac{1}{4}(2-\psi)^2
  \,,
\end{equation}
and it doesn't depend on the value of the parameter b.  The current
density is always sub-Goldreich-Julian, except the rotation axis,
where it is equal to the GJ current density.

\begin{figure}
  \includegraphics[clip,width=\columnwidth]{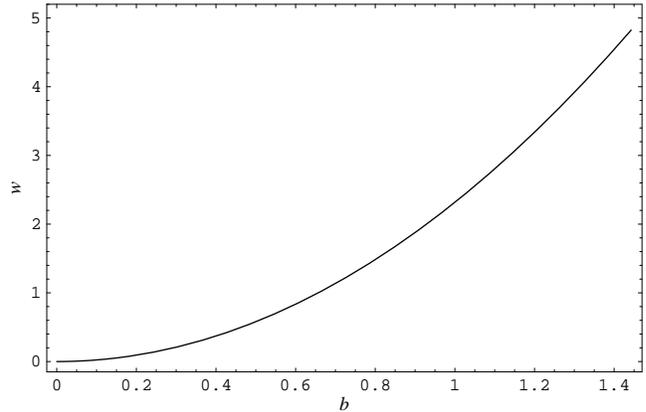}
  \caption{Spindown rate of an aligned rotator normalised to the
    spindown rate in the Michel solution for magnetosphere
    configurations with a constant current density across the cap as a
    function of the current density in the polar cap (parameter $b$).}
  \label{fig:a=0_w}
\end{figure}

The normalised spindown rate for the considered case has simple
quadratic dependence on the current density
\begin{equation}
  \label{eq:w_a_eq_0}
  \Const{w} = 6  (\log4 -1) b^2
  \,.
\end{equation}
This dependence is shown in Fig.~\ref{fig:a=0_w}.  The energy losses in
configuration with a constant current density could not be higher than
$\approx{}4.82$ of the energy losses in the corresponding Michel
solution.

The case $b=1$ is worth to mention separately, as it is ``the most
natural'' state for the space charge limited particle flow, for which
the current density at the surface on the NS is equal to the
corotational GJ current density.  In
Figs.~\ref{fig:a=0_v},~\ref{fig:a=0_beta},~\ref{fig:a=0_j} the lines
corresponding to this case are labelled with ``3''.  The maximum
potential drop for the configuration with the current density
distribution being equal to the corotational GJ current density is
$\Delta{}\Max{V}=0.386$ and the angular velocity of the open field
lines varies from $1$ at the rotation axis to $2$ at the polar cap
boundary.

%

\subsubsection{Configurations with the smallest potential drops}

%
\begin{figure}
  \includegraphics[clip,width=\columnwidth]{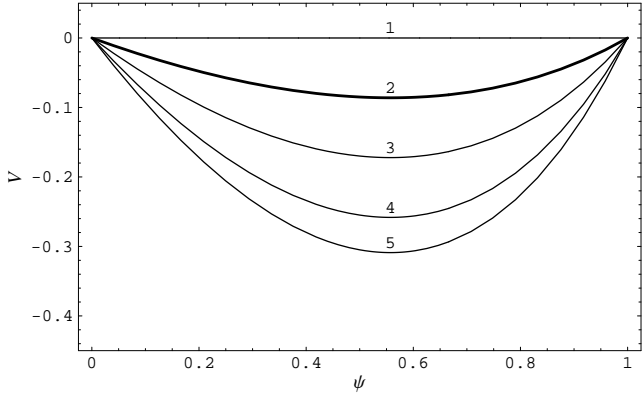}
  \caption{Electric potential in the polar cap of pulsar as a function
    of the normalised flux function $\psi$ for magnetosphere
    configurations with the smallest potential drop across the cap.
    In all cases $V_0$ is set to zero.  Numbered lines correspond to
    the following values of $a$: 1 --- $a=-1$ (Michel's solution); 2 ---
    $a=0$ (solution with a constant current density); 3 --- $a=1$; 4 ---
    $a=2$; 5 --- $a=1/(log4-1)$.}
  \label{fig:opt_v}
\end{figure}

%
\begin{figure}
  \includegraphics[clip,width=\columnwidth]{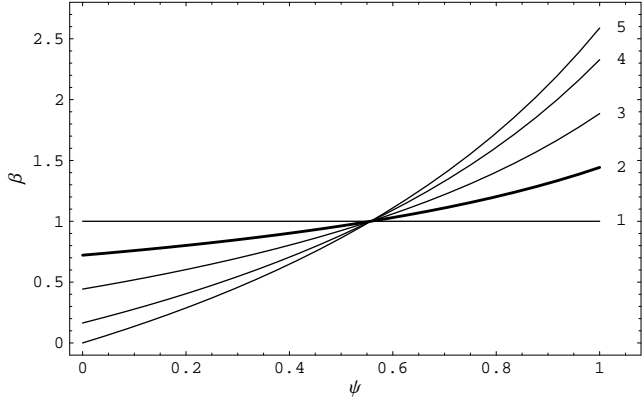}
  \caption{Normalised angular velocity of the open magnetic field
    lines as a function of the normalised flux function $\psi$ for
    magnetosphere configurations with the smallest potential drops
    across the cap. Labelling of the curves is the same as in
    Fig.~\ref{fig:opt_v}.}
  \label{fig:opt_beta}
\end{figure}

As the next example we consider the case when the maximum potential
drop across the polar cap for a fixed value of either $a$ or $b$ is
minimal.  The points corresponding to such values of parameters are
show in
Figs.~\ref{fig:Vmax_Contours},~\ref{fig:j/jGJ_Min},~\ref{fig:W} by the
thick grey line.  Equation for this line in the plane $(a,b)$ is
derived in Appendix~\ref{sec:App-maximum-potent-drop},
equation~(\ref{eq:DVb_is0}).  In some sense this is an optimal
configuration for the cascade zone, because for a fixed value of the
current density at some magnetic field line such configuration
requires the smallest potential drop among the other admitted
configurations.  The accelerating potential for the considered class
of configurations is
\begin{equation}
  \label{eq:V_ab_opt}
  \Opt{V}(\psi) = V_0 - (a+1) 
  \left\{
    \psi +\log\left[ \left( 1-\frac{\psi}{2} \right)^{\frac{1}{\log2}}\right]
  \right\}
  \,.
\end{equation}
The potential is shown as a function of $\psi$ in Fig.~\ref{fig:opt_v}
for several particular cases assuming for the sake of simplicity zero
potential drop at the polar cap boundary.  The potential has always a
minimum at the point
\begin{equation}
  \label{eq:psi_min__ab_opt}
  \Opt{\Min{\psi}} = 2-\frac{1}{\log 2} \simeq 0.557
  \,,
\end{equation}
the position of this minimum does not depend of the values of $a,b$.
The minimum value of the maximal potential drop across the cap,
$\min(\Max{\Delta{}V})=0$, is achieved at the left end of the grey
line, at the point $(a=-1,b=1)$ corresponding to the Michel solution.
The maximum potential drop across the gap for this class of
configurations, $\max(\Max{\Delta{}V})=0.309$, is achieved at the
right end of the grey line, at the point $(a=1/(\log4-1),b=0)$.

%
\begin{figure}
  \includegraphics[clip,width=\columnwidth]{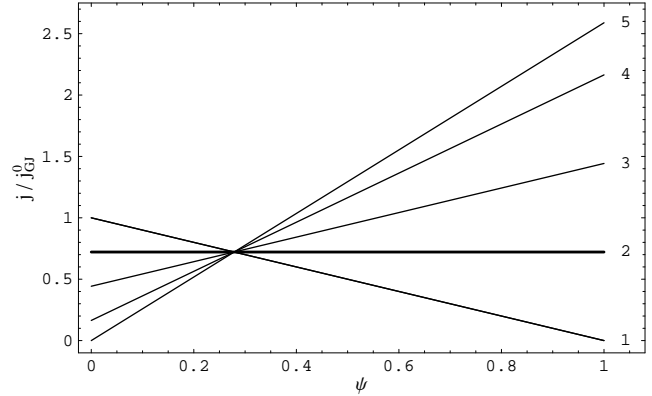}
  \caption{Current density as a function of the normalised flux
    function $\psi$ for magnetosphere configurations with the smallest
    potential drops across the cap. Labelling of the curves is the
    same as in Fig.~\ref{fig:opt_v}.}
  \label{fig:opt_j}
\end{figure}

%
\begin{figure}
  \includegraphics[clip,width=\columnwidth]{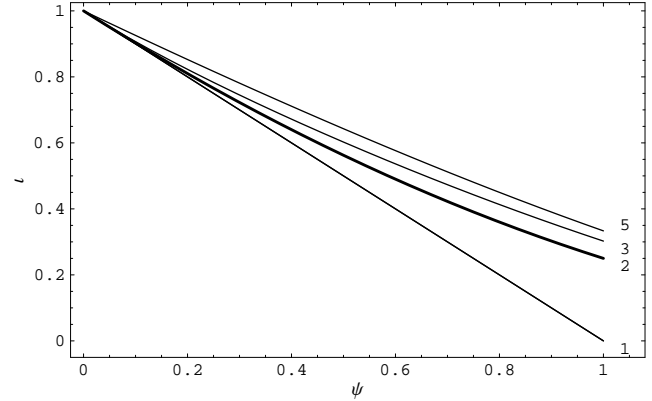}
  \caption{Ratio of the actual current density to the GJ current
    density $\iota$ as a function of the normalised flux function
    $\psi$ for magnetosphere configurations with the smallest
    potential drops across the cap. Labelling of the curves is the
    same as in Fig.\ref{fig:opt_v}.}
  \label{fig:opt_j2gj}
\end{figure}

The angular velocity of the open field lines is
\begin{equation}
  \label{eq:Beta_ab_opt}
  \Opt{\beta}(\psi) = \frac{a+1}{(2-\psi )\log2}-a
\end{equation}
Distribution of $\Opt{\beta}(\psi)$ is shown in
Fig.~\ref{fig:opt_beta}.  Before the minimum point $\Opt{\Min{\psi}}$
$\beta$ is not greater than 1, after that point $\beta$ is not smaller
than 1.  With increasing of the maximum potential drop the variation
of the angular velocity across the polar cap becomes larger.

The current density distribution in the considered case has the form
\begin{equation}
  \label{eq:j__ab_opt}
  \Opt{j}(\psi) = a \psi + \frac{a (1-\log4) + 1}{\log4} 
  \,.
\end{equation}
That distributions is shown in Fig.~\ref{fig:opt_j}. All curves pass
through the point $\hat{\psi}=1-1/\log4$, where the current density is
$\Opt{j}(\hat{\psi})=\GJNS{j}/\log4$.

The current density distribution in terms of the Goldreich-Julian
current density is
\begin{equation}
  \label{eq:j2jGJ_opt}
  \Opt{\iota}(\psi) = 
  \frac{(\psi -2)^2 \left[ (\psi -1)a\log4 + a + 1 \right]}%
  {\psi{}(4-\psi )a\log4 + 4a(1-\log4) + 4}
  \,,
\end{equation}
It monotonically decreases from 1 at the rotation axis to its minimum
value at the PC boundary.  This minimum value is in the range
$[0,1/3]$, the lowest value corresponds to the left end of the grey
curve (the Michel solution), the upper value corresponds to the right
end of the grey line.  The outflow is sub-GJ everywhere except the
rotation axis.

\begin{figure}
  \includegraphics[clip,width=\columnwidth]{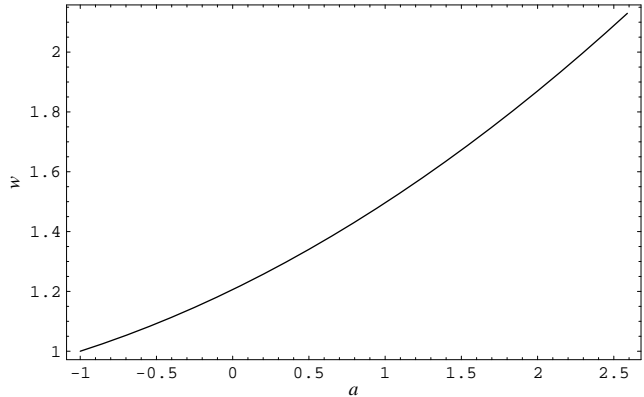}
  \caption{Spindown rate of an aligned rotator normalised to the
    spindown rate in the Michel solution for magnetosphere
    configurations with the smallest potential drops across the polar
    cap as a function of the parameter $a$.}
  \label{fig:opt_w}
\end{figure}

The normalised spindown rate for the considered case is
\begin{multline}
  \label{eq:w_opt}
  \Opt{w} = \frac{1}{2 \log^{2}2}
  \left\{
    \left[ 2\log^{2}2 - 3(1-\log2) \right] a^2 - 
  \right.\\
  \left.
    - 3(2-3\log2)a - 3(1-2\log2)
  \right\}
  \,.
\end{multline}
It is shown as a function of the parameter $a$ in
Fig.~\ref{fig:opt_w}. We see that for the considered configuration the
spindown rate as a function of the parameter $a$ increases more slowly
than the spindown rate for configurations with a constant current
density as a function of $b$, cf. Fig.\ref{fig:a=0_w}.  The energy
losses could not be higher than $\approx{}2.13$ of the energy losses
in the corresponding Michel solution.

%

\section{Discussion}
\label{sec:Discussion}

The main aim of this paper was to study the range of admitted current
density distributions in force-free magnetosphere of an aligned
rotator.  Taking into account that this subject was not studied in
details before, the linear model used in this work is in our opinion
an adequate approach to the problem.  The knowledge of the
magnetosphere behaviour in response to different potential drops in
the polar cap would be very useful for future modelling of
non-stationary polar cap cascades.  This formalism could be used as a
tool allowing to judge quickly whether a particular model of the polar
cap cascades is compatible with the force-free magnetosphere or not.
It may also give a clue how the magnetosphere would respond to a
particular current density distributions obtained at some step in
course of numerical solution.  Although the analytical model presented
here needs to be refined in numerical simulations, the presence of
some analytical relations should be very handy for numerical cascade
modelling.

We have considered here a simple case when the current density in the
polar cap of pulsar is a linear function of the magnetic flux.
However, the generalisation of the model to the case of a more
complicated shape of the current density distribution is
straightforward.  One should proceed with the steps described at the
beginning of the Section~\ref{sec:Equation_for_V} for the desired form
of the current density distribution.  The resulting equation for the
electric potential will be an ordinary differential equation, and
numerical solution of such equation for any given current density will
not be a problem.

The main conclusion we would like to draw from the presented results
is that even for rather moderate potential drop in the acceleration
zone the current density distribution can deviate significantly from
the ``canonical'' Michel distribution, which is basically preserved
for dipole geometry if all field lines corotates with the NS (see
\PapI).  In particular, a magnetosphere configuration with the
constant current density at the level of 73 per cents of the
Goldreich-Julian current density at the NS surface would require a
potential drop in the acceleration zone of the order of 10 per cents
of the vacuum potential drop.  For time-dependent cascades this may be
realised even for young pulsars.  We should note however that for
young pulsars the potential drop of the order of 10 per cents of the
vacuum drop could cause overheating of the polar cap by the particles
accelerated toward the NS
\citep[e.g.][]{Harding/Muslimov:heating_1::2001,Harding/Muslimov:heating_2::2002}.
In that sense such potential drop may be too large for young pulsars.
On the other hand, without knowledge of the dynamics of non-stationary
cascades it is in our opinion too preliminary to exclude the
possibility of such configurations for young pulsars, as in
non-stationary regime the heating of the cap may be not so strong as
in stationary cascades \citep[see][]{Levinson05}.

We used split-monopole approximation for the poloidal current density
distribution in the magnetosphere, which would produce an accurate
results only for configurations with very small size of the corotating
zone, less than the size of the light cylinder
$x_0\ll{}1/\beta(\PsiL)$, see Fig.~\ref{fig:Magnetosphere}.  For the
(most interesting) case when these sizes are comparable, results
obtained in this work could be considered only as a zero approximation
to the real problem.  We should point out an important modification
introduced by the dipole geometry of the magnetic field.  For the
dipole geometry there will be some magnetic field lines which bent to
the equatorial plane at the light cylinder.  For these field lines the second term%
\footnote{which in the cylindrical coordinates $(\varpi,\phi,z)$ is
  $2\beta\pd{}_\varpi\Psi$}
on the r.h.s. of equation~(\ref{eq:CondAtLC}) will be negative, and in
order to get positive current density along these lines a more steep
dependence of $\beta$ on $\psi$ would be necessary.  As a result the
potential drop in the dipole geometry would be higher than it is
obtained in our approximation.  Figuratively speaking, in our model we
could correct only for the decrease of the electric current density
toward the polar cap boundary present in the Michel solution.  On the
other hand, if $\beta(\PsiL)>1$, the size of the corotating zone can
be less as well as greater%
\footnote{for the closed magnetic field lines the angular velocity is
  $\Omega$ and they are still inside \emph{their} light cylinder; the
  adjustment of the angular velocity occurs in the current sheet,
  which is for sure a non-force-free domain.}
then the light cylinder radius at the last open field line.  In the
latter case there should be less magnetic field lines which bent
toward the equatorial plane, than in the first case, cf.
Fig.~\ref{fig:Magnetosphere} cases \textbf{I} and \textbf{II}.  Hence,
the correction introduced by the dipolar field geometry for some
subset of our solutions would be non-monotonic as the size of the
corotating zone $x_0$ increases.  So, there is still possibility that a
moderate potential drop could allow a large variety of the current
densities, although this issue needs careful investigation.

In this paper we ignored the electrodynamics of the polar cap zone.
Although without a theory of time-depending cascades we cannot put
more limits on the electric potential than the one we used in
Section~\ref{sec:Beat(psi)}, there is an additional limitation coming
from the basic electrodynamics.  Namely, the accelerating potential
near conducting walls, which the current sheet at the polar cap edge
is believed to be, should approach zero.  However, we could speculate
that there is a thin non-force-free zone at the edge of the polar cap
where the adjustment of the potential happens.  In other words, the
return current and the actually nearly equipotential region may occupy
not the whole non-force-free zone.  Because of this that limitation
would not eventually restrict our solutions.

At the end we would like to discuss briefly the issue with the pulsar
breaking index.  If the inner pulsar magnetosphere is force-free, then
the spindown rate of an aligned pulsar as a function of angular
velocity will deviate from the power law $W\propto\Omega^4$ if the
size of the corotating zone or/and the distribution $\beta(\psi)$
change with time.  The assumption that these ``parameters'' are time
dependent seems to us to be natural, because with the ageing of pulsar
the conditions in the polar cap cascade zone change and the
magnetosphere should adjust to these new conditions.  In the frame of
our model we could make some simple estimations how the braking index
of pulsar is affected by the changes of these two ``parameters''.  

As an example we consider the case when the pulsar magnetosphere
evolves through a set of configurations with a constant current
density.  The spindown rate for such configurations is
\begin{equation}
  \label{eq:W_x0_b___a_0}
  W \propto \Omega^4 \PsiL^2 b^2 \sim \Omega^4 x_0^{-2} b^2 
  \,,
\end{equation}
here we estimate \PsiL{} assuming dipole field in the corotation zone.
If $b$ and/or $x_0$ are functions of time, the spindown rate will be
different from the spindown of a dipole in vacuum.  If at some moment
the dependence of the size of the corotating zone and the current
density on $\Omega$ could be approximated as $x_0 \propto \Omega^\xi$
and $b_0 \propto \Omega^\zeta$ correspondingly, the braking index of
pulsar measured at that time is
\begin{equation}
  \label{eq:nbreak__a_0}
  n = 3 - 2\xi + 2\zeta
  \,.
\end{equation}
We see, that the deviation of the braking index from the ``canonical''
value being equal to 3 would be by a factor of two larger than the
dependence of $b$ and $x_0$ on $\Omega$.  The braking index could be
smaller as well as larger than 3, depending on the sign of the
expression $\zeta-\xi$.  For instance, if in an old pulsar the
potential drop increases and, as the consequence, the current density
decreases, $\zeta$ is positive, and the braking index could be
\emph{greater} than 3.  We note, that there are evidences for such
values of the braking index for old pulsars
\citep{Arzoumanian/Chernoff/Cordes:2002}.  On the other hand, if $x_0$
decreases with the time, as it was proposed in \PapI{} for young
pulsars, the braking index would be less than 3.  However even in this
simple picture different trends may be possible.  In reality the
evolution of the pulsar magnetosphere will be more complicated and
more steep as well as more gradual dependence of the braking index on
the current density would be possible.  This would results in a rather
wide range of possible values of pulsar braking index.

\section*{Acknowledgments}

I acknowledge J.~Arons, J.~Gil, Yu.~Lyubarsky and G.~Melikidze for
fruitful discussions.  I would like to thank J.Arons for useful
suggestions to the draft version of the article.  This work was
partially supported by the Russian grants N.Sh.-5218.2006.2,
RNP.2.1.1.5940, N.Sh.-10181.2006.2 and by the Israel-US Binational
Science Foundation

\bibliographystyle{mn2e} 
\bibliography{psreq}

\begin{thebibliography}{}

\bibitem[\protect\citeauthoryear{{Al'Ber}, {Krotova} \& {{E}idman}}{{Al'Ber}
  et~al.}{1975}]{AlBer/Krotova:1975}
{Al'Ber} Y.~I.,  {Krotova} Z.~N.,    {{E}idman} V.~Y.,  1975, Astrophysics, 11,
  189

\bibitem[\protect\citeauthoryear{{Arons}}{{Arons}}{1979}]{Arons79}
{Arons} J.,  1979, Space Science Reviews, 24, 437

\bibitem[\protect\citeauthoryear{{Arzoumanian}, {Chernoff} \&
  {Cordes}}{{Arzoumanian} et~al.}{2002}]{Arzoumanian/Chernoff/Cordes:2002}
{Arzoumanian} Z.,  {Chernoff} D.~F.,    {Cordes} J.~M.,  2002, \apj, 568, 289

\bibitem[\protect\citeauthoryear{{Beskin}, {Gurevich} \& {Istomin}}{{Beskin}
  et~al.}{1993}]{GurevichBeskinIstomin_Book}
{Beskin} V.,  {Gurevich} A.,    {Istomin} Y.,  1993, {Physics of the Pulsar
  Magnetosphere}.
Cambridge University Press

\bibitem[\protect\citeauthoryear{{Beskin}}{{Beskin}}{2005}]{BeskinBook}
{Beskin} V.~S.,  2005, "Osesimmetrichnye stacionarnye techeniya v astrofizike"
  (Axisymmetric stationary flows in astrophysics).
Moscow, Fizmatlit, \textit{in russian}

\bibitem[\protect\citeauthoryear{{Blandford} \& {Znajek}}{{Blandford} \&
  {Znajek}}{1977}]{Blandford/Znajek77}
{Blandford} R.~D.,  {Znajek} R.~L.,  1977, \mnras, 179, 433

\bibitem[\protect\citeauthoryear{{Bucciantini}, {Thompson}, {Arons}, {Quataert}
  \& {Del Zanna}}{{Bucciantini} et~al.}{2006}]{Bucciantini06}
{Bucciantini} N.,  {Thompson} T.~A.,  {Arons} J.,  {Quataert} E.,    {Del
  Zanna} L.,  2006, \mnras, 368, 1717

\bibitem[\protect\citeauthoryear{{Contopoulos}}{{Contopoulos}}{2005}]{Contopou%
los05}
{Contopoulos} I.,  2005, \aap, 442, 579

\bibitem[\protect\citeauthoryear{{Contopoulos}, {Kazanas} \&
  {Fendt}}{{Contopoulos} et~al.}{1999}]{CKF}
{Contopoulos} I.,  {Kazanas} D.,    {Fendt} C.,  1999, \apj, 511, 351

\bibitem[\protect\citeauthoryear{{Fawley}}{{Fawley}}{1978}]{Fawley/PhDT:1978}
{Fawley} W.~M.,  1978, PhD thesis, AA(California Univ., Berkeley.)

\bibitem[\protect\citeauthoryear{{Goldreich} \& {Julian}}{{Goldreich} \&
  {Julian}}{1969}]{GJ}
{Goldreich} P.,  {Julian} W.~H.,  1969, \apj, 157, 869

\bibitem[\protect\citeauthoryear{{Goodwin}, {Mestel}, {Mestel} \&
  {Wright}}{{Goodwin} et~al.}{2004}]{Goodwin/04}
{Goodwin} S.~P.,  {Mestel} J.,  {Mestel} L.,    {Wright} G.~A.~E.,  2004,
  \mnras, 349, 213

\bibitem[\protect\citeauthoryear{Gruzinov}{Gruzinov}{2005}]{Gruzinov:PSR}
Gruzinov A.,  2005, Phys.Rev.Lett., 94, 021101

\bibitem[\protect\citeauthoryear{{Harding} \& {Muslimov}}{{Harding} \&
  {Muslimov}}{2001}]{Harding/Muslimov:heating_1::2001}
{Harding} A.~K.,  {Muslimov} A.~G.,  2001, \apj, 556, 987

\bibitem[\protect\citeauthoryear{{Harding} \& {Muslimov}}{{Harding} \&
  {Muslimov}}{2002}]{Harding/Muslimov:heating_2::2002}
{Harding} A.~K.,  {Muslimov} A.~G.,  2002, \apj, 568, 862

\bibitem[\protect\citeauthoryear{{Komissarov}}{{Komissarov}}{2006}]{Komissarov%
06}
{Komissarov} S.~S.,  2006, \mnras, 367, 19

\bibitem[\protect\citeauthoryear{{Levinson}, {Melrose}, {Judge} \&
  {Luo}}{{Levinson} et~al.}{2005}]{Levinson05}
{Levinson} A.,  {Melrose} D.,  {Judge} A.,    {Luo} Q.,  2005, \apj, 631, 456

\bibitem[\protect\citeauthoryear{{Lyubarskij}}{{Lyubarskij}}{1992}]{Lyubar92}
{Lyubarskij} Y.~E.,  1992, \aap, 261, 544

\bibitem[\protect\citeauthoryear{{McKinney}}{{McKinney}}{2006}]{McKinney:NS:06}
{McKinney} J.~C.,  2006, \mnras, 368, L30

\bibitem[\protect\citeauthoryear{{Michel}}{{Michel}}{1973a}]{Michel73:b}
{Michel} F.~C.,  1973a, \apj, 180, 207

\bibitem[\protect\citeauthoryear{{Michel}}{{Michel}}{1973b}]{Michel73}
{Michel} F.~C.,  1973b, \apjl, 180, L133

\bibitem[\protect\citeauthoryear{{Michel}}{{Michel}}{1991}]{MichelBook}
{Michel} F.~C.,  1991, {Theory of neutron star magnetospheres}.
Chicago, IL, University of Chicago Press, 1991, 533 p.

\bibitem[\protect\citeauthoryear{{Okamoto}}{{Okamoto}}{1974}]{Okamoto74}
{Okamoto} I.,  1974, \mnras, 167, 457

\bibitem[\protect\citeauthoryear{{Ruderman} \& {Sutherland}}{{Ruderman} \&
  {Sutherland}}{1975}]{Ruderman/Sutherland75}
{Ruderman} M.~A.,  {Sutherland} P.~G.,  1975, \apj, 196, 51

\bibitem[\protect\citeauthoryear{{Scharlemann}, {Arons} \&
  {Fawley}}{{Scharlemann} et~al.}{1978}]{Arons/Scharlemann/78}
{Scharlemann} E.~T.,  {Arons} J.,    {Fawley} W.~M.,  1978, \apj, 222, 297

\bibitem[\protect\citeauthoryear{{Scharlemann} \& {Wagoner}}{{Scharlemann} \&
  {Wagoner}}{1973}]{ScharlemannWagoner73}
{Scharlemann} E.~T.,  {Wagoner} R.~V.,  1973, \apj, 182, 951

\bibitem[\protect\citeauthoryear{{Spitkovsky}}{{Spitkovsky}}{2006}]{Spitkovsky%
:incl:06}
{Spitkovsky} A.,  2006, \apjl, 648, L51

\bibitem[\protect\citeauthoryear{{Timokhin}}{{Timokhin}}{2006}]{Timokhin2006:M%
NRAS1}
{Timokhin} A.~N.,  2006, \mnras, 368, 1055

\bibitem[\protect\citeauthoryear{{Timokhin}}{{Timokhin}}{2007}]{Timokhin::PSRE%
Q2/2007}
{Timokhin} A.~N.,  2007, \apss, in press, astro-ph/0607165

\end{thebibliography}

\appendix
\section[]{Admitted current density}
\label{sec:App--admitt-curr-dens}

%
\begin{figure}
  \includegraphics[clip,width=\columnwidth]{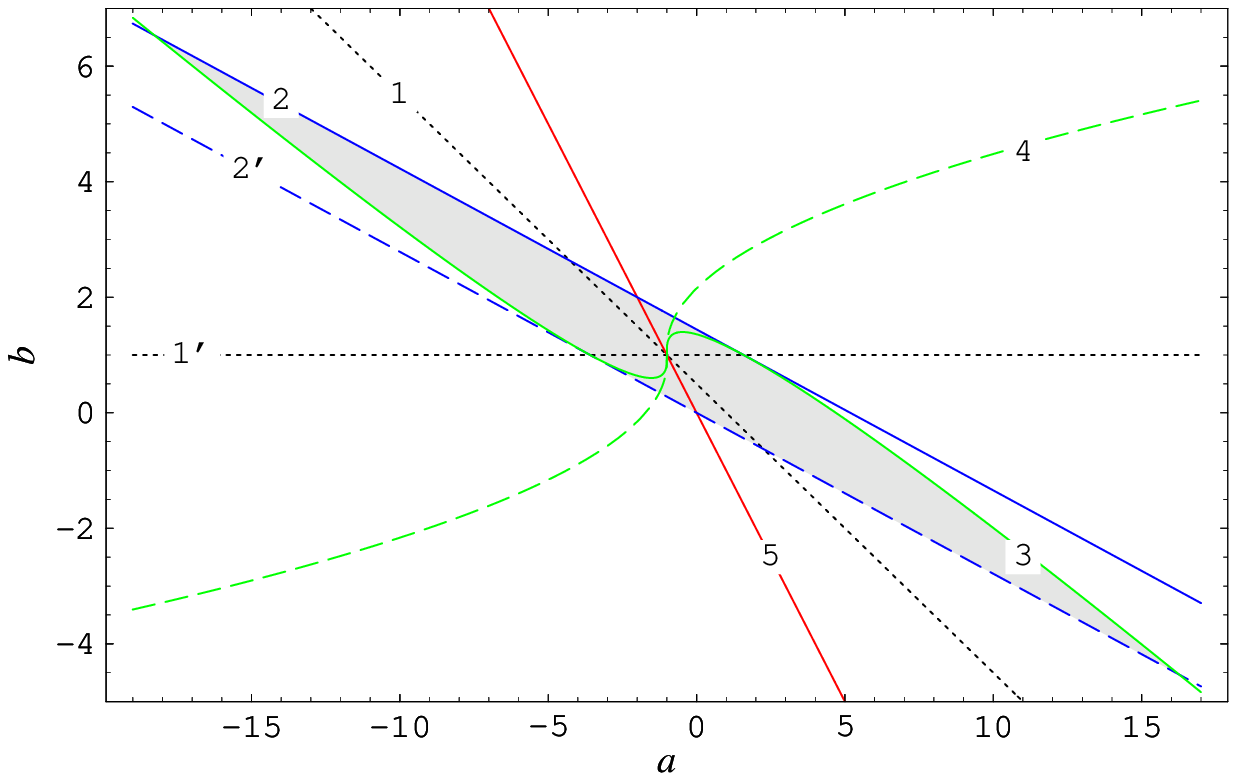}
  \caption{Value of parameters $a$ and $b$ admitted by the requirement
    that the maximum potential drop across the polar cap is less than
    the vacuum potential drop.  Dotted lines $\bmath{1}$ and
    $\bmath{1'}$ show the boundary of the region where the potential
    has an extremum in the polar cap.  Line $\bmath{2}$ corresponds to
    $\DVba=1$, and $\bmath{2'}$ -- to $\DVba=-1$.  Line $\bmath{3}$
    show points when $|\DVbe|=1$, and $\bmath{4}$ -- $|\DVae|=1$.
    Line $\bmath{2'}$ corresponds to $b=-a$. The resulting admitted
    region is shown by the grey colour.  See text for the explanations.}
  \label{fig:ab_range_full}
\end{figure}

The second derivative of $V(\psi)$ in respect to $\psi$ is
\begin{equation}
  \label{eq:d2V_dpsi2}
  V''(\psi) = 2\,\frac{a+b}{(2-\psi )^2}
  \,.
\end{equation}
For fixed $a$ and $b$ the second derivative never changes the sign,
hence, $V(\psi)$ has a single extremum.  If $a+b>0$, $V(\psi)$ has a
single minimum; in the plane $(a,b)$ shown in
Fig.~\ref{fig:ab_range_full} these points lie left to the line
$\bmath{5}$.  If $a+b<0$, $V(\psi)$ has a single maximum, such points
lies to the right of the line $\bmath{5}$ in
Fig.~\ref{fig:ab_range_full}.  The potential reaches its extremum
value at the point
\begin{equation}
  \label{eq:psi_min}
  \PsiEx = 2\,\frac{1-b}{a+1}
  \,,
\end{equation}
This point lies in the interval $[0,1]$ if 
\begin{equation}
  \begin{split}
    \label{eq:b_psi_min_inside_pc}
    1 \ge b \ge \frac{1-a}{2},& \quad \mbox{ for } a > -1\,;\\
    1 \le b \le \frac{1-a}{2},& \quad \mbox{ for } a < -1\,.
  \end{split}
\end{equation}
In Fig.~\ref{fig:ab_range_full} points where these conditions are
satisfied lie between the lines $\bmath{1}$ and $\bmath{1'}$, in the
region where the angle between the lines is acute.  The line
$\bmath{1}$ corresponds to the values of $a,b$ when the extremum of
the potential is reached at the polar cap boundary.  For $a,b$ lying
on the line $\bmath{1'}$ the potential reaches the extremum at the
rotation axis.

If $\PsiEx$ is outside of the interval $[0,1]$, $V(\psi)$ is a
monotone function of $\psi$ in the polar cap of pulsar and the maximum
potential drop is the potential drop between the edge of the polar cap
and the rotation axis.  In that case the
condition~(\ref{eq:DeltaV_less_1}) takes the form
\begin{equation}
  \label{eq:DeltaV_less_1_psi_min_outside_pc}
  |\DVba|\le 1
  \,,
\end{equation}
where $\DVba\equiv{}V(1)-V(0)$ is the potential drop between the the
boundary of the polar cap and the rotation axis.  In terms of the
parameters $a,b$ it can be written as
\begin{equation}
  \label{eq:b_DeltaV_less_1_psi_min_outside_pc}
  a\,\frac{1-\log4}{\log4}
  \le b \le
  \frac{1}{\log2}+ a\,\frac{1-\log4}{\log4}
  \,.
\end{equation}
In Fig.~\ref{fig:ab_range_full} points satisfying this condition lie
between the lines $\bmath{2}$ and $\bmath{2'}$.  In the region, where
the angle between the lines $\bmath{1}$ and $\bmath{1'}$ is abuse
(here the point $\PsiEx$ is outside of the interval $[0,1]$) the lines
$\bmath{2}$ and $\bmath{2'}$ sets the boundaries for admitted values
of the parameters $a$ and $b$.

If the electric potential reaches its extremum value inside the polar
cap, the maximum potential drop is achieved either between the extremum
point and the edge of the polar cap, or between the extremum point and
the rotation axis.  In that case the condition~(\ref{eq:DeltaV_less_1})
takes the form
\begin{equation}
  \label{eq:DeltaV_less_1_psi_min_inside_pc}
  \max( |\DVba|,|\DVae|, |\DVbe| ) \le 1
  \,,
\end{equation}
where $\DVae\equiv{}V(0)-V(\PsiEx)$ is the potential drop between the
rotation axis and the point $\PsiEx$, and
$\DVbe\equiv{}V(1)-V(\PsiEx)$ is the potential drop between the polar
cap boundary and the point $\PsiEx$.  Expression for the extremum
value of $V$ is non-linear in respect to $a,b$:
\begin{equation}
  \label{eq:Vex}
  V(\PsiEx) = V_0 - 1 + a + 2b - 
  2(a+b)\log\left[ \frac{2(a+b)}{1+a} \right]
  \,,
\end{equation}
and condition~(\ref{eq:b_psi_min_inside_pc}) in terms of $a,b$ should
be evaluated numerically.  For a fixed $a$ the derivatives of $\DVae$
and $\DVbe$ in respect to $b$ are
\begin{eqnarray}
  \label{eq:dDeltaVae_db}
  \frac{d\,\DVae}{d\,b} & = & 2\log\left(\frac{a+b}{1+a}\right)\\
  \label{eq:dDeltaVbe_db}
  \frac{d\,\DVbe}{d\,b} & = & 2\log\left(\frac{a+b}{1+a}\right) + \log4  
  \,.
\end{eqnarray}
If $\PsiEx\in[0,1]$ and condition~(\ref{eq:b_psi_min_inside_pc}) is
fulfilled, then $d\,\DVbe/d\,b$ is positive and $d\,\DVae/d\,b$ is
negative.  So, for a fixed $a$ when $\PsiEx\in[0,1]$ $\DVae$ decreases
and $\DVbe$ increases with increasing of $b$.

In Fig.~\ref{fig:ab_range_full} the line $\bmath{3}$ represents points
where $|\DVbe|=1$, and the line $\bmath{4}$ -- the points where
$|\DVae|=1$.  To the right of the line $\bmath{5}$ $V(\psi)$ has a
minimum, and the lines $\bmath{3}$ and $\bmath{4}$ represent points
where $\DVbe=1$ and $\DVae=1$ correspondingly.  To the left of the
line $\bmath{5}$ $V(\psi)$ has a maximum, and here the lines
$\bmath{3}$ and $\bmath{4}$ correspond to $\DVbe=-1$ and $\DVae=-1$.
The line $\bmath{4}$ lies always outside of the region between lines
$\bmath{1}$ and $\bmath{1'}$, and, hence, the absolute value of the
potential drop between the extremum point and the rotation axis
$|\DVae|$ never achieves the vacuum potential drop when the extremum
point is inside the interval $(0,1)$.

$\DVbe$ increases with increasing of $b$.  So, below the curve
$\bmath{3}$ to the right of the line $\bmath{5}$, and above the curve
$\bmath{3}$ to the left of the line $\bmath{5}$ $|\DVae|$ is less
than~1.  Hence, when $\PsiEx\in[0,1]$ ($a,b$ are in the area between
the lines $\bmath{1}$ and $\bmath{1'}$), the admitted values of $a$
and $b$ are limited by the lines $\bmath{3}$ and $\bmath{2'}$ to the
left from the line $\bmath{5}$, and by the lines $\bmath{2}$ and
$\bmath{3}$ to the right from the line $\bmath{5}$.

Combining all discussed restrictions we get the region of the admitted
values of parameters $a$ and $b$, which is shown in
Fig.~\ref{fig:ab_range_full} by the grey area.

\section[]{The maximum potential drop across the polar cap}
\label{sec:App-maximum-potent-drop}

%
\begin{figure}
  \includegraphics[clip,width=\columnwidth]{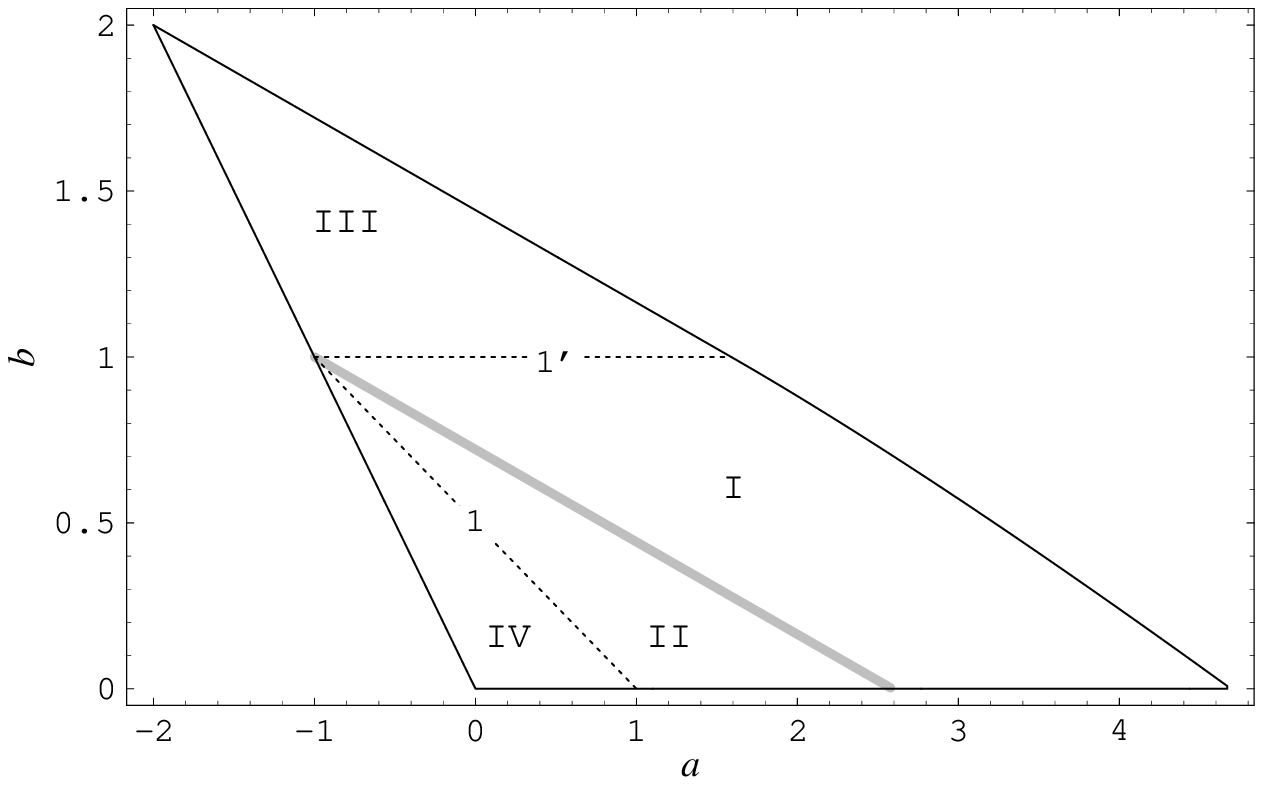}
  \caption{Region of admitted values of the parameters $a$ and $b$
    corresponding to current densities being of the same sign as the
    Goldreich-Julian current density.  The dotted line $\bmath{1}$ shows
    points where $\DVbe=0$, the line $\bmath{1'}$ -- points where
    $\DVae=0$.  The thick grey shows points where $|\DVae|=|\DVbe|$.
    In the region I -- $\Max{\Delta{}V}=\DVbe$, in the region II --
    $\Max{\Delta{}V}=\DVae$, in the regions II and IV --
    $\Max{\Delta{}V}=|\DVba|$. See explanations in the text.}
  \label{fig:ab_range}
\end{figure}

In Fig.~\ref{fig:ab_range} the dotted lines $\bmath{1}$ and
$\bmath{1'}$ limit the region%
\footnote{regions I and II together}
in the parameter space $(a,b)$ where the potential in the polar cap is
a non-monotone function of $\psi$ and it has a minimum at some point
$\PsiEx\in{}(0,1)$, see Appendix~\ref{sec:App--admitt-curr-dens}.  In
the regions III and IV the potential $V(\psi)$ is a monotone function
of $\psi$.

The potential drop $\DVae\equiv{}V(0)-V(\PsiEx)$ between the rotation
axis and the point $\PsiEx$ where $V(\psi)$ achieves its minimum
value, for a fixed $a$ decreases with increasing of $b$.  For $a,b$ at
the line $\bmath{1'}$ $\PsiEx$ is at the rotation axis, and $\DVae=0$
there.  The potential drop $\DVbe\equiv{}V(1)-V(\PsiEx)$ between the
polar cap boundary and the point $\PsiEx$ for a fixed $a$ increases
with increasing of $b$.  For $a,b$ on the line $\bmath{1}$ $\PsiEx$ is
at the polar cap boundary and $\DVbe=0$.  So, $\DVae$ increases in the
direction from the line $\bmath{1'}$ to the line $\bmath{1}$, and
$\DVbe$ increases in the direction from the line $\bmath{1}$ to the
line $\bmath{1'}$.

Along some line between lines $\bmath{1'}$ and $\bmath{1}$ the
potential drops $\DVae$ and $\DVbe$ become equal.  This means also
that the potential drop $\DVba\equiv{}V(1)-V(0)$ between the polar cap
edge and the rotation axis is zero there.  Equation for this line is
obtained easily from the requirement $\DVba=0$:
\begin{equation}
  \label{eq:DVb_is0}
  b = \frac{1}{\log4} + a\frac{1-\log4}{\log4}
  \,.
\end{equation}
This line is shown in Fig.~\ref{fig:ab_range} by the thick grey line.
Above the grey line in Fig.~\ref{fig:ab_range} $\DVbe>\DVae$, and
below it $\DVbe<\DVae$.  Hence, the line given by
eq.~(\ref{eq:DVb_is0}) is the line where the maximum potential drop
across the polar cap achieves its minimum value for fixed $a$ or $b$.

Taking all this into account we conclude, that the maximum potential
drop across the polar cap  $\Max{\Delta{}V}$ is equal to the following
potential drops: in the region I -- to $\DVbe$, in the
region II -- to $\DVae$, in regions III and IV -- to $|\DVae|$.

\section[]{Energy of the electromagnetic field in split-monopole
  configuration}
\label{sec:energy-electr-field}

The energy density of the electromagnetic field in the magnetosphere
is
\begin{equation}
  \label{eq:w_General}
  w = \frac{1}{8\pi}(E^2+\Pol{B}^2+B_\phi^2)
  \,.
\end{equation}
For the split monopole solution when $\Psi=\PsiL(1-\cos\theta)$ the
non-zero components of the electric and magnetic fields are
\begin{eqnarray}
  \label{eq:E_B_components_monopol_E}
  E_\theta &=& - \frac{\mu}{\RLC^3} \PsiL\beta \frac{\sin\theta}{r}\\
  \label{eq:E_B_components_monopol_Br}
  B_r    &=& \frac{\mu}{\RLC^3} \frac{\PsiL}{r^2}\\
  \label{eq:E_B_components_monopol_Bf}
  B_\phi &=& - \frac{\mu}{\RLC^3} \PsiL\beta \frac{\sin\theta}{r}
  \,.
\end{eqnarray}
So, the electric field is equal to the toroidal magnetic field,
$E_\theta=B_\phi$.  The total energy of the magnetosphere is then
\begin{equation}
  \label{eq:W__E_Br_monopol}
  \mathcal{W} =\frac{1}{8\pi}
  \int \left( B_r^2 + 2 E_\theta^2 \right)\,dV
\end{equation}
On the other hand, the energy losses are
\begin{equation}
  \label{eq:Spindown_monopol}
  W = 
  \int_{4\pi} c\, \frac{\left[\vec{E}\times\vec{B}\right]_r}{4\pi}\,d\tilde{\Omega} =
  \int_{4\pi} c\, \frac{E_\theta^2}{4\pi}\,d\tilde{\Omega}
  \,,
\end{equation}
where $\tilde{\Omega}$ is a solid angle.  Using
equations~(\ref{eq:Spindown_monopol}),~(\ref{eq:E_B_components_monopol_Br})
we can rewrite the expression~(\ref{eq:W__E_Br_monopol}) for the total
energy of electromagnetic filed as
\begin{multline}
  \label{eq:W__Spindown}
  \mathcal{W} = 
  \frac{\Wmd}{c} \times\\
  \left[
    \frac{\PsiL^2 \RLC^2}{2} \left(\frac{1}{\RNS} - \frac{1}{R}\right) + W(R-\RNS)
  \right]
  \,,
\end{multline}
where $R$ is the size of the magnetosphere (where it dynamics is
determined by the central object).  The first term represents the
total energy of the poloidal magnetic field and it is the same for all
configurations with different potential drops.  The second term, being
the sum of the energies of the electric field and the toroidal
component of the magnetic field is different for different values of
the accelerating potential. It is directly proportional to the
energy losses in a particular configuration.

\label{lastpage}
\end{document}